\documentclass[journal]{vgtc}     
    \onlineid{0}
    \vgtccategory{Research}
    \vgtcpapertype{application/design study}
    
    \title{Towards Intelligent Augmented Reality (iAR): A Taxonomy of Context, an Architecture for iAR, and an Empirical Study}
    
    \author{
        \authororcid{Shakiba Davari}{0000-0003-3128-1979}
            \thanks{e-mail: sdavari@vt.edu} \\
         \and \authororcid{Daniel Stover}{0009-0004-5358-8423}\\
         \and \authororcid{Alexander Giovannelli}{0000-0002-0265-8143}\\
         \and \authororcid{Cory Ilo}{0009-0001-1208-8260}\\
        \and \authororcid{Doug. A. Bowman}{0000-0003-0491-5067}
    }
    \affiliation{\scriptsize Center for Human-Computer Interaction \\ Department of Computer Science, Virginia Tech, Blacksburg, VA, USA}
    
    \abstract{
    Recent advancements in Augmented Reality (AR) research have highlighted the critical role of context awareness in enhancing interface effectiveness and user experience. This underscores the need for intelligent AR (iAR) interfaces that dynamically adapt across various contexts to provide optimal experiences.
    In this paper, we \textbf{(a)} propose a comprehensive framework for context-aware inference and adaptation in iAR, \textbf{(b)} introduce a taxonomy that describes context through quantifiable input data, and \textbf{(c)} present an architecture that outlines the implementation of our proposed framework and taxonomy within iAR.
    Additionally, we present an empirical AR experiment to observe user behavior and record user performance, context, and user-specified adaptations to the AR interfaces within a context-switching scenario. We \textbf{(d)} explore the nuanced relationships between context and user adaptations in this scenario and discuss the significance of our framework in identifying these patterns.
    This experiment emphasizes the significance of context-awareness in iAR and provides a preliminary training dataset for this specific Scenario. 
    }
    \renewcommand{\manuscriptnotetxt}{}

\usepackage{xargs}      
\usepackage{soul}       
\usepackage{color}      
\usepackage{xspace}     
\usepackage{xpunctuate} 
\usepackage{tabu}                      

\usepackage{changepage}   

\newcommandx{\guest}[3][1=]
    {\setulcolor{lightorange}{\ul{#1}} \textcolor{lightorange} 
    {[\textbf{#2:} #3]}}
\newcommandx{\shaq}[2][1=] 
    {\setulcolor{purple}{\ul{#1}} \textcolor{purple}   
    {[\textbf{Shakiba:} #2]}}  

\newcommand{\badge}[2]{\colorbox{#1}{\textcolor{white}{\textsc{#2}}}}
\newcommand{\headerBadge}[2]{
  \vspace{-15px}                  
  \hspace*{\fill}                 
  \badge{#1}{#2}                  
  \vspace{4px}\linebreak\noindent 
}
\newcommandx{\due}[1]{\headerBadge{purple}{Due: #1}}  

\usepackage{graphicx} 
\usepackage{subfigure}

\graphicspath{{files/}{figs/}{./}} 
\newcommand{\figContextSettingSUI}{
\begin{figure*}[t!]
    \vspace*{-3em}
    \centering
    \hspace*{-5em}
    \includegraphics[width=.9\linewidth]{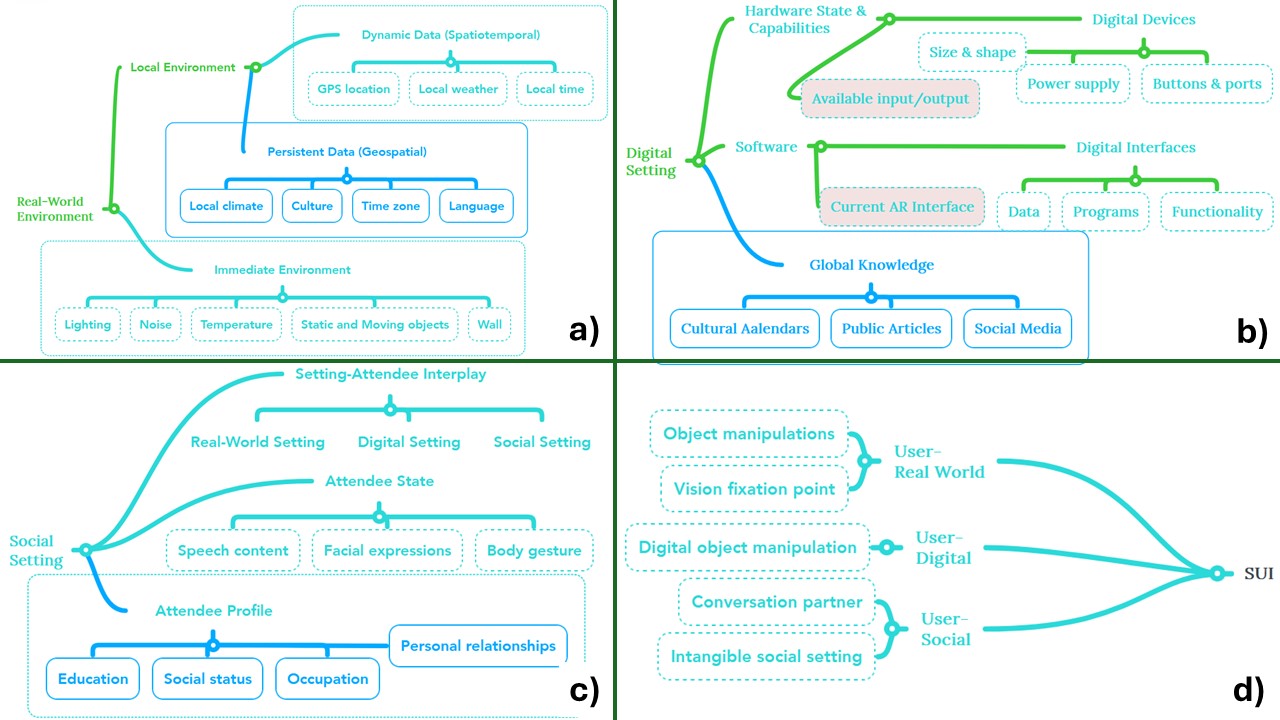}
    \vspace*{-9pt}
    \caption{Persistent and Transient context components present:
    \textbf{(a)}: the structure and state of physical items and phenomena in the user's local and immediate environment. 
    \textbf{(b)}: public/global information and the hardware and firmware state and capabilities of the AR and available digital devices.
    Transient context components present: 
    \textbf{(c)}: accessible information from the profile and state of physical/virtual attendees in the current context and their interactions with the real-world/digital setting. 
    \textbf{(d)}: the connections and interactions of the user's profile and state with their social, real-world, and digital setting.}
    \label{fig:contextAll}
\vspace*{-1.5em}
\end{figure*}
}

\newcommand{\figtaxoUser}{
\begin{figure}[htb]
    \centering
    \vspace*{-12pt}
    \includegraphics[width=0.5\textwidth]{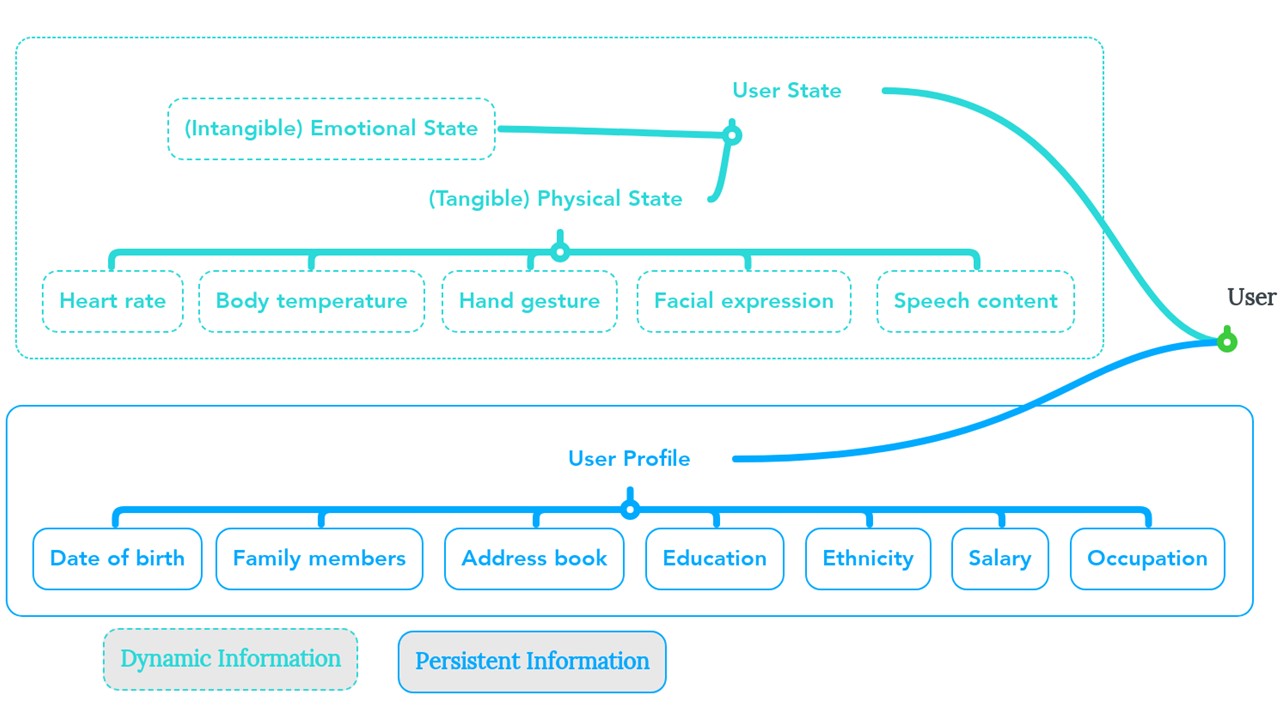}
    \vspace*{-22pt}
    \caption{Persistent and transient context components provide the iAR with information about the user profile and state.}
    \vspace*{-18pt}
    \label{fig:taxoUser}
\end{figure}
}

\newcommand{\figCiabox}{
\begin{figure}[htb]
    \centering
    \includegraphics[width=0.5\textwidth]{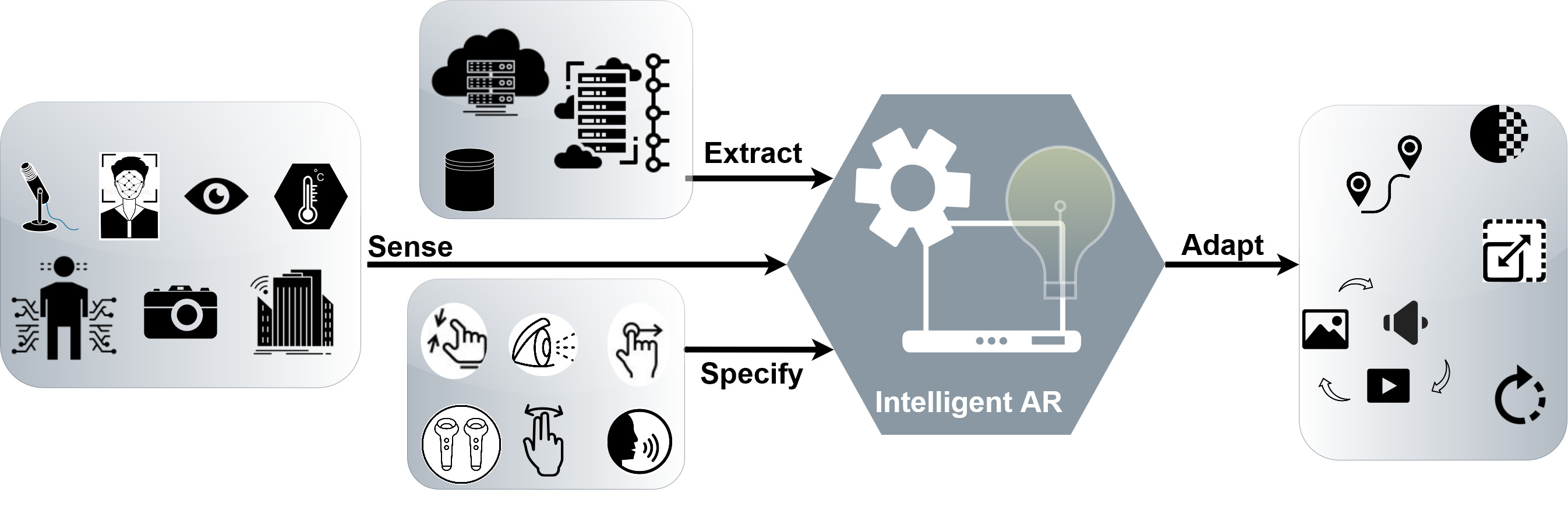}
    \caption{Intelligent AR interfaces adapt the interface based on quantifiable input data from the user, sensors, and personal or publicly available data storage and databases.}
    \label{fig:cia-box}
\end{figure}
}

\newcommand{\figFramework}{
\begin{figure}[htb]
    \centering
    \vspace*{-10pt}
    \includegraphics[width=0.5\textwidth]{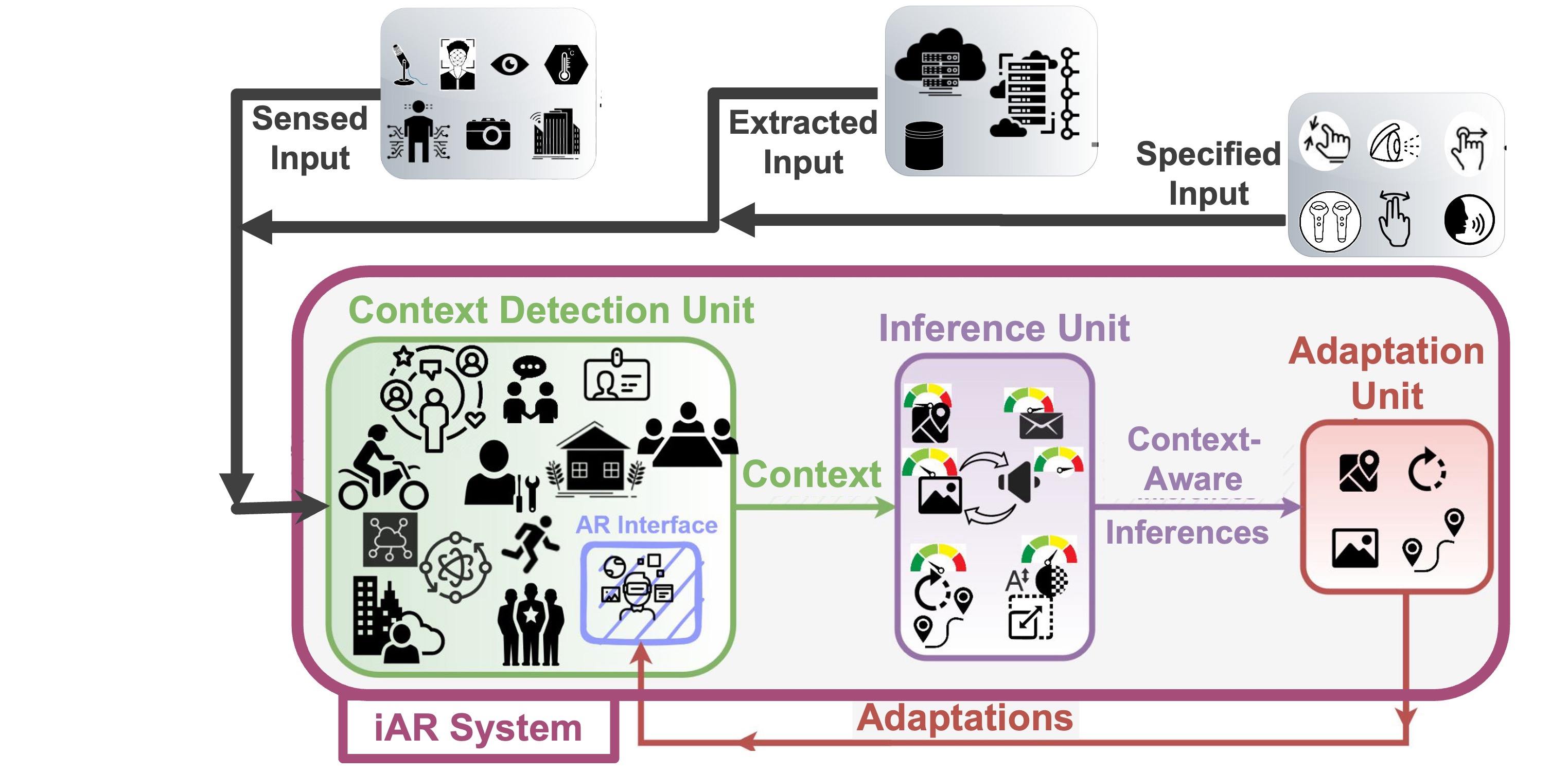}
    \vspace*{-15pt}
    \caption{Our framework for Context-aware Inference and Adaptation using our taxonomy of Context.}
    \vspace*{-10pt}
    \label{fig:framework}
\end{figure}
}

\newcommand{\figArch}{
\begin{figure}[htb]
    \centering
    \includegraphics[width=0.5\textwidth]{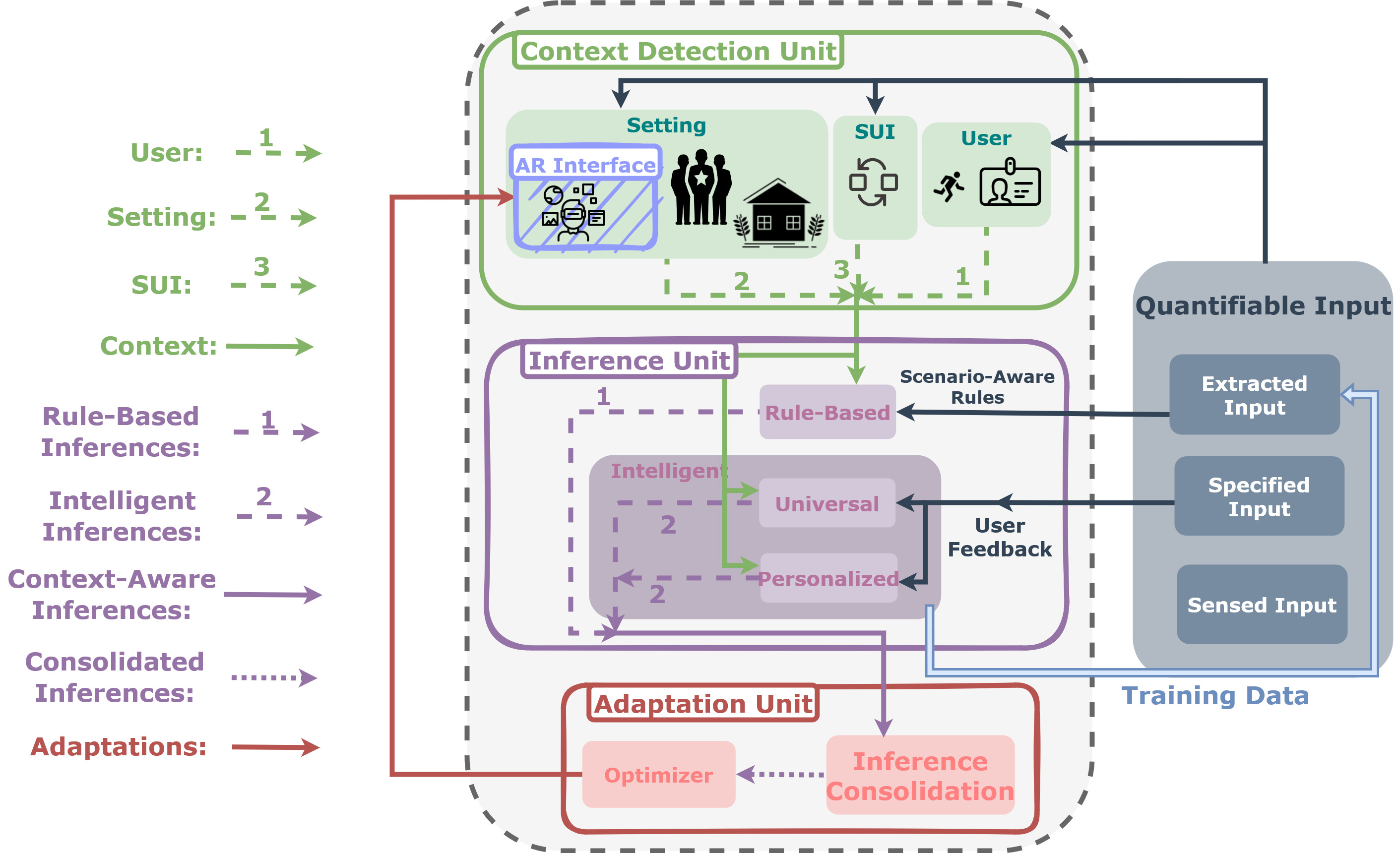}

    \vspace*{-5pt}
    \caption{The architecture of intelligent AR systems.}
    \vspace*{-14pt}
    \label{fig:frameworkDetailed}
    \label{fig:architecture}
\end{figure}
}
\newcommand{\figScreenshots}{
\begin{figure}[htb]
    \vspace*{-14pt}
    \centering
    \subfigure[]{\includegraphics[width=0.24\textwidth]{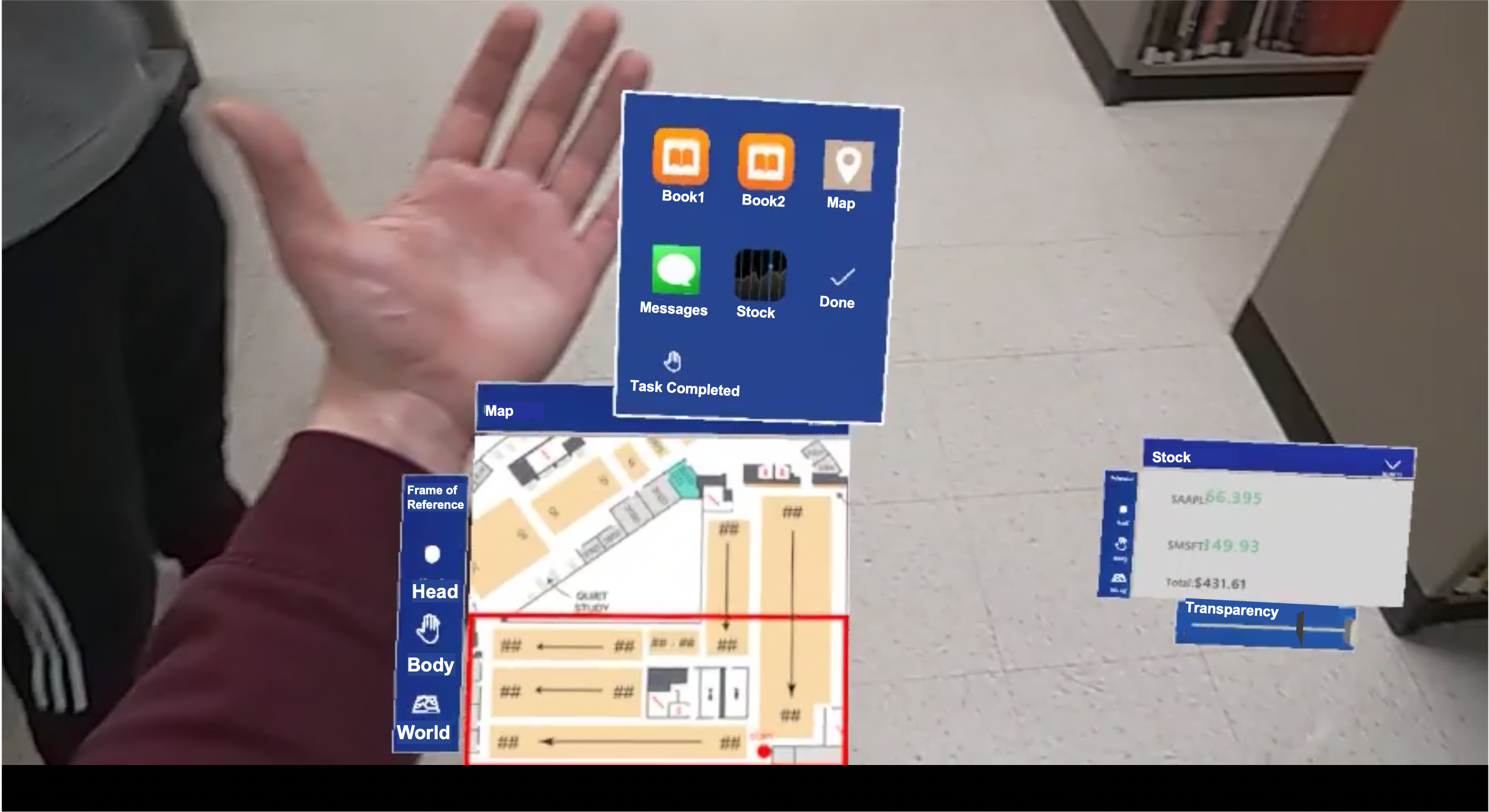}\label{fig:screenshots-a}}
    \subfigure[]{\includegraphics[width=0.24\textwidth]{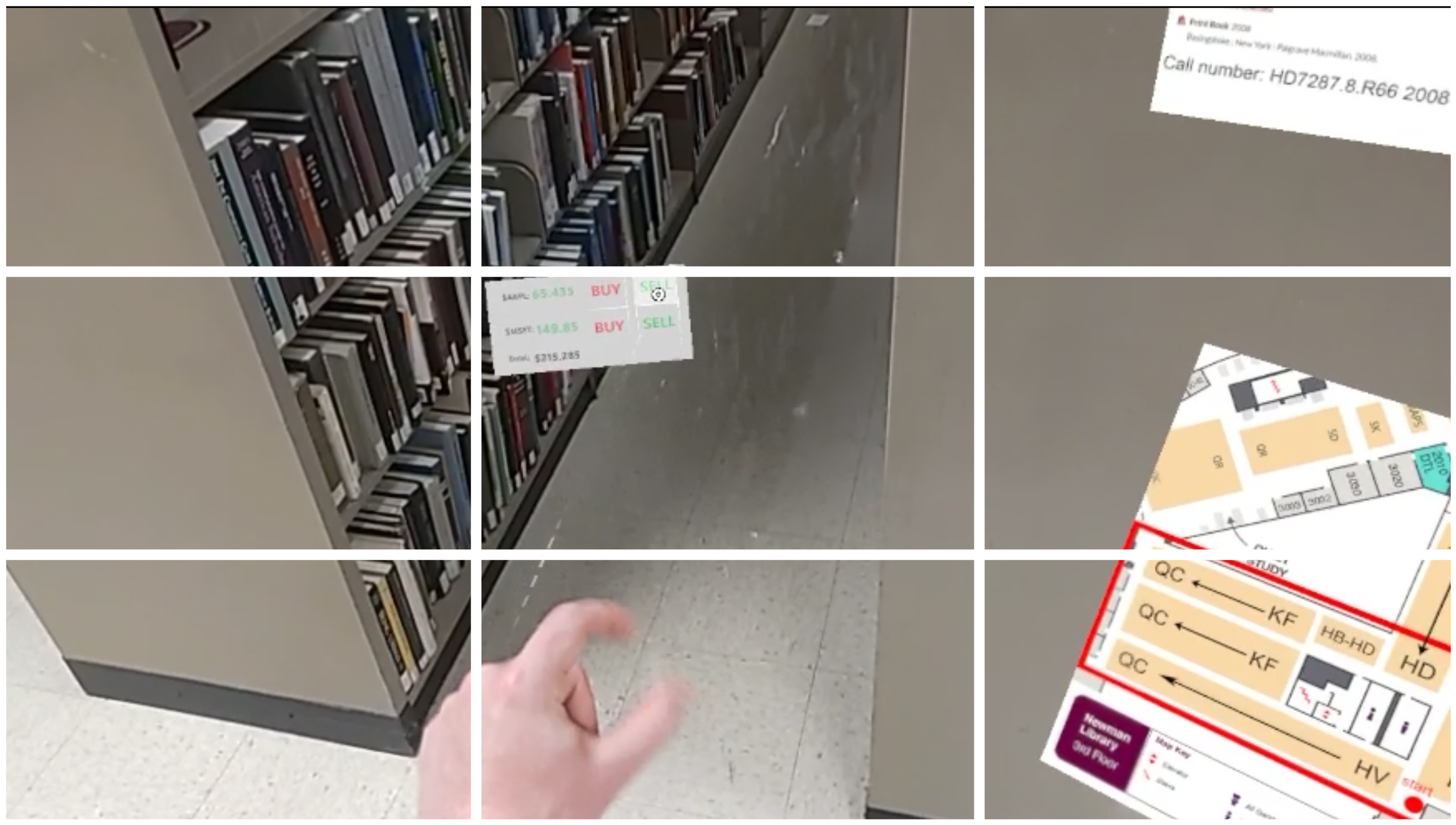}\label{fig:screenshots-b}}
        \vspace*{-10pt}
    \caption{   (a) User making adaptations to the apps in the Edit Mode  
                (b) Head-fixed sectors indicated on a screenshot of the user's view.}
    \label{fig:screenshots}
        \vspace*{-12pt}
\end{figure}
}

\usepackage{mathptmx}

\begin{document}
\firstsection{Introduction}
\maketitle
Augmented Reality (AR) integrates digital content into the real world, enabling simultaneous, continuous, and effortless access to multiple pieces of information anytime and anywhere \cite{feiner2002augmented, weiser1997coming, matthews2007designing, weiser1999computer}.
However, distractions, real-world occlusion, and information overload in AR can challenge users' performance, perception, and awareness.
Unlike physical devices, AR's boundless potential for ubiquitously displaying information within the user's environment complicates the design of a universally optimal interface.
Whether for entertainment or information access in offices, streets, meetings, or emergencies, users engage with AR in diverse contexts, each with varying requirements, constraints, and preferences that impact its effectiveness.
Optimal AR must identify the appropriate AR content, presentation, and interaction within a specific context, highlighting the importance of context awareness in AR interface design \cite{davari2022dc, grubert2016towards}.
\par
We distinguish between context-aware AR and intelligent AR (iAR).\\
\textbf{Context-aware AR} adapts based on \textit{scenario-specific design principles} using a \textit{limited range of contextual components}. 
However, AR design complexity stems from the interdependence and interaction of various contextual components. Thus, by overlooking certain contextual components, the effectiveness of these context-aware interfaces remains constrained.
\textbf{Intelligent AR (iAR)}, aims to predict the \textit{impact of various adaptations} based on the current state of \textit{all relevant contextual components} that may influence AR effectiveness without relying on predefined design principles.
\par
While prior research proposed taxonomies for context, many of them overlook components essential to AR interface effectiveness or involve implicit elements without a clear framework for detecting and quantifying them within an iAR system.
Towards designing iAR interfaces, in this work, we explore three \textbf{Research Questions (RQs)}:
    \\
    \textbf{\textit{RQ1.}} \textit{How can context be described to an iAR through explicit and quantifiable components?}\\
From the literature, we investigate prior conceptualizations and characterizations of context, along with the factors that influence AR effectiveness. Through an iterative refinement process, brainstorming, and a thorough literature review, we develop a comprehensive taxonomy of quantifiable contextual components that impact AR effectiveness.
    \\
    \textbf{\textit{RQ2.}} \textit{How can iAR systems utilize our taxonomy of explicit context to infer implicit information about the effectiveness of their design?}\\
The complexity of context awareness is not just detecting the contextual components but intelligently inferring how the specific combination of all these components should be used to adapt the interface.
For instance, while overlaying AR content on other individuals’ faces might be highly intrusive during a conversation \cite{davari2022validating}, it could be deemed acceptable or even beneficial when studying in a crowded library.
Towards providing the optimal AR, we propose an iAR architecture to integrate contextual components and learn to optimize AR effectiveness by inferring the impact of various adaptations.

\textbf{\textit{RQ3.}} \textit{How do users adapt their AR interface in a context-switching scenario?}\\
Designing an effective iAR necessitates understanding the intricate relationship between the context and the user-preferred adaptations to various AR design dimensions, requiring training data involving user adaptations across diverse contexts.
\par
To address these \textbf{RQs}, we make the following contributions.
In \Cref{se:taxonomy}, we propose a comprehensive taxonomy of explicit and quantifiable contextual components for iAR.
To provide a road map towards designing iAR, \Cref{se:architecture} presents a framework utilizing our taxonomy and proposes a potential architecture for context-aware inference and adaptation in iAR. 
In \Cref{se:experiment4}, we present an empirical study observing context, user performance, and user-specified adaptations to an AR interface in a context-switching scenario. We analyze this data in \Cref{se:results} to explore how users adapt the interface based on contextual factors. The results highlight the importance of implicit inferences for optimizing AR effectiveness and enhancing user experience in iAR.
\section{Related Work}
In today's life, the growing dependence on personal computing devices such as mobile phones for decision-making and task performance has led to challenges, such as disruptions to social interactions \cite{sohn2008diary, dearman2008examination, church2014large, abeele2016effect, hall2014put}. This has driven Ubiquitous computing's goal of seamlessly embedding near-constant, yet unobtrusive, digital information into daily life \cite{weiser1997coming, weiser1999computer}. Research indicates that by integrating the information into the real world (RW), AR can enhance efficiency and memory retention and reduce such challenges \cite{valimont2002effectiveness, davari2020occlusion}. However, AR effectiveness depends on presenting the right information at the right time and form \cite{biener2021extended, feiner1999arissues, grubert2016towards}. This work focuses on developing iAR systems that enhance user experience and efficiency by ensuring timely and appropriate information delivery in any context.
\par
Various studies, such as ARWin and ARBrowse, have demonstrated the benefits of AR for providing on-demand information, assistance, and entertainment, spanning from domestic use to workplace collaboration and productivity \cite{knierim2019exploring, di2003arwin, langlotz2014next}. However, the effectiveness of AR interfaces is influenced by factors such as information overload, visual clutter, obtrusiveness, and distractions, all of which can negatively affect situational awareness, cognitive load, and performance \cite{krevelen2010survey, bengler2006augmented, endsley1988design, grubert2010extended, stoltz2017augmented}. For instance, in social contexts, poorly designed AR interfaces can obstruct communication by occluding facial expressions, leading to social isolation and privacy concerns \cite{gugenheimer2018facedisplay, gugenheimer2019challenges, adolphs1999social, frith2009role}. This has led to research on AR design principles for non-invasive AR and view management that provide socially relevant information while maintaining visual focus on interlocutors \cite{hollerer2001user, orlosky2015halo, davari2022validating}.
\par
The RW Spatial setting and occlusion also significantly impact AR effectiveness \cite{livingston2003resolving, li2019gaze, cutting1995perceiving, drascic1996perceptual}. Extensive research has proposed AR design principles to address occlusion between RW and virtual objects \cite{lepetit2000semi, shah2012occlusion, wloka1995resolving, berger1997resolving, matthews2007designing}. These principles often emphasize view management techniques that ensure non-occlusive yet visible placement of virtual content \cite{bell2001view, azuma2003evaluating, grasset2012image, makita2009view}, as well as peripheral placement approaches \cite{cadiz2001sideshow, maglio2000tradeoffs, lu2020glanceable}. Other strategies prioritize the real world, activating virtual content only when necessary through user-triggered adaptations, such as transparency, layout, or level of detail (LoD) \cite{ens2014personal, ens2015spatial}. This work examines the underexplored impact of RW spatial settings and occlusions on AR effectiveness, focusing on user-specified AR adaptations in context-switching scenarios.
\par
An effective AR interface must include considerations such as positioning, transparency, and LoD for timely and appropriate information presentation in a given context.
Pervasive AR envisions ``continuous, universal, and omnipresent" integration of AR content into various daily tasks \cite{grubert2016towards}.
Non-adaptive AR interfaces, which fail to adjust to user context, risk presenting information inappropriately.
Achieving Pervasive AR requires a deep understanding of the user’s changing context, task, and environment to inform interface design.
AR devices, equipped with sensors, offer unique opportunities for context detection, as seen in research assisting visually impaired users \cite{ed2021socialsensemaking}, highlighting AR's potential to meet everyday information needs.
\par
Context-aware interfaces, such as ARWin and HoloDoc, propose design principles to optimize content presentation and enhance user experiences in specialized contexts using contextual data like user attention, spatial settings, lighting, and environmental cues \cite{langlotz2014adaptivebrowser, vertegaal2003attentive, white2009InteractionAP}. However, an interface suitable in one scenario may be ineffective in another, as user priorities shift across contexts, requiring the interface to adapt accordingly \cite{harrison2008lean, diverdi2004level}. For instance, while providing relevant information to the user's conversation can enhance user experience \cite{sohn2008diary, davari2022validating},  the user's cognitive load may require low levels of information if they are engaged in a high-level collaborative task \cite{rhodes1998wimp, billinghurst1999wearable}.
User-triggered, context-aware interfaces allow manual selection of contextually-customized AR content presentation, offering predictability and control but increasing user effort \cite{roy2019automation, rivu2020stare}.
Recent research has focused on automatic adaptations to AR content without manual intervention, demonstrating the integration of unobtrusive context-aware AR into daily life \cite{lindlbauer2019context, langlotz2014next, cheng2021semanticadapt}.
\par
To enable Pervasive AR, the interface must automatically detect and respond to contextual changes. However, current context-aware interfaces are limited in their adaptation scope, detect only a narrow range of contextual components, such as fatigue or social interactions, and use predefined design principles tailored to specific scenarios \cite{chang2018drowsy, davari2022validating}. This work investigates iAR interfaces capable of optimal adaptations in unfamiliar contexts without prior knowledge. 
\par
Enabling iAR requires a framework that supports all potential AR adaptations, the detection and representation of the context, and its delivery to the decision-making process \cite{perera2013context, dey2001conceptual}.
Previous work identifies a design space of the potential adaptations to an AR interface \cite{davari2024designspace}.
Various approaches have been proposed for representing context, such as the 5 Ws (Who, What, Where, When, Why), or using categories like location, identity, time, and activity \cite{perera2013context, abowd1999context, schilit1994context}. Others study specific sub-contexts such as physical environment \cite{grasset2012image, cheng2021semanticadapt}, objects \cite{white2009InteractionAP, brandon2020objectaware}, background \cite{blair2001backgroundaware}, and depth perception \cite{feng2021bi}.
However, these taxonomies often overlook components essential to AR interface effectiveness, incorporate implicit elements that cannot be automatically detected or quantified by an iAR system, or lack clear frameworks for identifying implicit information essential for AR design.
In this work, we reviewed existing research on context characterization and taxonomies and, through iterative refinement, identified key contextual information that influences AR performance. We propose a comprehensive taxonomy of quantifiable contextual components and a framework for iAR systems that use these components to infer the impact of various AR adaptations and make optimal adjustments in real-time.

\section{Context-Aware Inference and Adaptation}\label{se:architecture}
The context dictates the user's specific limitations, wants, and needs, influencing their preferences for the AR interface at each time.
To properly adapt, an iAR must use its input data to identify these user preferences in each context (\Cref{fig:cia-box}).
We classify the iAR input into three types:
\figCiabox
      
      \vspace*{3pt}
      
      \noindent  
\textbf{\textit{Specified input}} are quantifiable information \textit{provided manually} by the user.
iAR coherently aims to limit reliance on this input type.
That is because these inputs are limited to specific human-understandable data and prone to human error.
They can also be slow and inaccurate, distract the user, and increase their physical and mental load, which contradicts these interfaces' objective of increasing efficiency and effectiveness.
      
      \vspace*{3pt}
      
      \noindent  
\textbf{\textit{Sensed input}} are raw or processed quantifiable information \textit{directly received} from sensors.
Sensed inputs from various sensors like a camera, microphone, accelerometer, gyroscope, and magnetometer provide information about the environment, the user, the device's pose, etc. 
This input type can also include processed sensor data for recognizing and tracking faces, objects, voice commands, gestures, and gaze.
      
      \vspace*{3pt}
      
      \noindent  
\textbf{\textit{Extracted input}} are quantifiable information \textit{retrieved from previously stored} data
on various personal or publicly available data storage and databases through in-network and web queries.
These data storage and databases are collections of information that were \textit{specified} or \textit{sensed} in the past.
For example, extracted inputs can provide information such as the GPS location and time of day.
\par
    To address \textbf{RQ2.}, we propose a framework to design iAR that receives context as their input, infers the user's preferences for the interface, and makes the proper adaptations accordingly (\Cref{fig:framework}).
    This section provides a taxonomy to describe the context within this framework and proposes an architecture demonstrating how we envision implementing our framework and taxonomy within iAR systems.
    \figFramework
    \par
    To inform the design and development of iAR, it is essential to describe the context comprehensively through data that can serve as input for these systems.
    This framework describes \textbf{\textit{Context}} as the knowledge derived from all the available \textit{contextual components} that can affect the AR interface's design dimensions.
    These \textit{\textbf{contextual components}} are considered as any piece of \textit{specified}, \textit{extracted}, or \textit{sensed} input representing \textit{dynamic} or \textit{persistent} contextual information.
    \par
    \textbf{\textit{Persistent contextual components}} are static or change infrequently and are quantified by \textit{extracted} inputs.
    Information about the user's education, local time zone, or family members is a few examples of persistent contextual components.
    While recording a change in these contextual components is through \textit{specified} or \textit{sensed} inputs, they will be stored on an internal or external data storage and \textit{extracted} when accessed for context detection.
    \par
    \textbf{\textit{Transient contextual components}} are prone to more frequent and unexpected changes and need to be quantified more frequently through \textit{sensed} or \textit{extracted} input, e.g., user's body gesture, speech content, environmental lighting.
    \par
    \Cref{se:taxonomy} provides a taxonomy, classifying contextual components that influence the AR interface into three main categories: \nameref{se:user}, \nameref{se:setting}, and \nameref{se:sui}.
    As previously discussed, most contextual components can be \textit{specified}. However, iAR aims to replace this input type when possible, especially for \textit{Transient contextual components}. 
    To reduce reliance on \textit{specified} inputs, this taxonomy identifies alternative \textit{extracted} and \textit{sensed} input types, for each contextual component.
    \par
    We define an \textbf{Inference} as an interface design dimension's level of the contingency on context (\Cref{fig:framework}).
    Since the AR interface is part of the virtual setting, an inference may correlate multiple interface design dimensions. 
    We refer to inferences that only correlate interface design dimensions as \textit{Design Inferences}, while \textit{Context-aware Inferences} refer to those involving one or a combination of multiple contextual components
    \par
    \textit{Inferences} indicate the potential impact of specific \textbf{Adaptations} on the performance of the interface.
    The inference unit receives the context, described in \Cref{se:taxonomy}, as its input and, using different methods and sources, deduces a collection of \textit{contextual inferences} in the present context.
    However, each design dimension can be involved in numerous inferences.
    Within a context, multiple inferences from different sources can provide duplicated or contradictory adaptations to a specific design dimension.
    To automatically adapt and optimize the interface's effectiveness, this framework proposes an \textit{Adaptation} process to integrate these inferences and identify the optimal non-conflicting set of adaptations for the interface.
    \par
    Given the multitude of contextual components and their potential variations, it is infeasible to replicate all possible contexts, much less to determine the optimal adaptation that the iAR system must apply for each unique context. 
    We define a \textit{\textbf{scenario}} as a collection of distinct contexts with at least one common contextual component.
    For example, a walking scenario encompasses contexts such as walking in the street, walking on a treadmill, playing soccer, and any other contexts in which the user's physical state is walking.
    \textit{Inferences} enables an iAR interface to generalize the known optimal adaptations for a specific context to other contexts within the same scenario.
    Additionally, \textit{Inferences} enable an iAR interface to deduce information such as \textit{User Task} and \textit{User Objective} that are influenced by multiple contextual components and cannot be captured through a single contextual component.
    While such information can be included within context through \textit{specified} contextual components, they are usually limited or inaccurate (e.g., in a task-switching scenario). 
    For example, imagine Sanaz working from home and contacting the IT specialist through non-work-related social media to resolve an issue with the company’s Slack channel.
    Without inferences, a specified \textit{User Task} contextual component may indicate she is ``working", leading her AR interface to close non-work related social media and interrupt her communication with the IT specialist.
    \textit{Inferences} enable her iAR to correlate her \textit{Location}, the \textit{Disconnected Slack}, and \textit{the IT specialist} to prevent disrupting their communication.
\subsection{Taxonomy of Context}\label{se:taxonomy}
    To design a taxonomy of explicit and quantifiable contextual components that influence the effectiveness of AR interfaces, we adopted a comprehensive and systematic methodology. 
    Initially, to avoid biases present in existing literature, we conducted multiple brainstorming sessions to enumerate all conceivable contextual components. During this process, we explored the XR design space and identified design elements that can adapt the content and presentation of an XR interface \cite{davari2024designspace}.
    We thoroughly evaluated the challenges and capabilities associated with these design dimensions and compiled a list of contextual components that could influence them.
    For instance, when considering the color theme as a design element, a pertinent challenge is that blue colors are not suitable for nighttime use and can cause eye fatigue. This underscores the importance of ``the time of day" as a crucial contextual component.
    As another example, the position, size, and opacity design elements of an AR object can result in the occlusion of important objects in RW. This emphasizes the vitality of contextual components providing information about the user's interplay with objects and people within the setting.
    \par
    To ensure the inclusion of all pertinent aspects from the literature, we reviewed well-established taxonomies for describing context and analyzed the contextual information utilized in various proposed context-aware AR interfaces.
    These steps enabled us to compile a comprehensive list of contextual information.
    Subsequently, we engaged in an iterative refinement process, continuously reassessing the relevance of each contextual component, ensuring a robust and effective taxonomy. 
    \par
    Abowd et al. establish location, identity, time, and activity as the main context categories \cite{perera2013context},
    Grubert et al., categorize context into user, physical environment, and social/cultural context; other research has focused on adaptations based on objects \cite{white2009InteractionAP, brandon2020objectaware}, background \cite{blair2001backgroundaware}, depth perception \cite{feng2021bi}, and physical environment \cite{grasset2012image, cheng2021semanticadapt} from the context.
    We identify three main categories of quantifiable contextual components that influence AR effectiveness either directly or indirectly through inferences (e.g., see the scenario of Sanaz working from home in \Cref{se:architecture}). While the Setting category provides information about the state of the world, excluding the user, the User category focuses solely on the user, and the Setting-User Interplay category provides information about the interactions between context components from these categories.
        The contextual components of each context category are classified based on their input type or the semantics of the information they describe.
        \subsubsection{User} \label{se:user}
        Who the user is, what they are currently doing, how they feel, and other information about the user themselves affect the type of information they need and their preferences, resources, and limitations for accessing and interacting with this information.
        For example, a chef may need less detailed instructions from a recipe, while a beginner requires the same recipe to be presented in more detail and include visualizations. 
        In our taxonomy, we classify the contextual components only about the user, isolated from their setting, as the user category of context.
        Based on the input type, we
        classify the contextual components within the User category into two subcategories: User Profile and User State (\Cref{fig:taxoUser}).
        \figtaxoUser
            \\
            \\
            \textbf{Persistent User Profile}\\ 
            Persistent information about users and their characteristics, such as their values and responsibilities, sociocultural factors, socioeconomic status, psychological and physiological abilities, and traits, can affect their interface.
            For example, a visually impaired person may require a different font size than a person with normal vision.
            \par
            \textit{Persistent User profile} provides established information about the user through \textit{persistent contextual components}. 
            These are contextual components such as meaningful dates, locations, and people, user's physical limitations, assets, properties, weekly schedule, salary, occupation, and responsibilities that can be \textit{extracted} from a user's digital profiles, health records, legal documents, bank accounts, etc.
            \par
            Due to their staticity and infrequent changes, user profile contextual components are \textit{extracted} inputs. 
            This information is specified or sensed if changed; however, when accessing input to detect the context they are extracted. 
            For example, a student specifies their demographic and education information and the classes they are taking on their digital profile.
            Inputs such as their class schedules and course syllabus \textit{extracted} from this digital profile and the learning management system allow their AR interface to provide the class material for them automatically.
            Similarly, inputs such as their level of education and customer preferences can be \textit{extracted} from this digital profile.
                  
      \vspace*{4pt}
      
      \noindent  
            \textbf{Transient User State}\\ 
            Adaptations to a user's interface can also be affected by transient factors such as their current physical activity, conversation topic, and mental and emotional state.
            \figContextSettingSUI
            For example, the user's emotional state can affect content adaptation, such as presenting the user's anxiety management tool during a panic attack.
            \textit{User state} provides transient information about the user through \textit{Transient contextual components} that change continually. 
            Contextual components like heart rate, body temperature, gestures, and speech content can be obtained using sensor data from an altimeter, 3-axis accelerometer, optical and temperature sensor, microphone, camera, gesture sensors, etc.
            \par
            Since the user state is subject to unpredictable and frequent changes, these components must be \textit{sensed}.
            However, with accuracy at stake, current technological limitations may require combining or replacing \textit{sensed inputs} with other contextual component types to remove or reduce reliance on specific sensors. 
            For instance, when proper sensors are unavailable, costly, or require high power consumption, a combination of \textit{specified} and \textit{extracted} inputs can be used as an alternative. 
        In such cases, some contextual components can be \textit{sensed} at specific intervals rather than constantly. 
        During these intervals, these contextual components will be \textit{extracted} from the previous measures.
        As another example, the user's emotional state can be \textit{specified}, directly \textit{sensed} from EEG sensors, or \textit{extracted} from the user's mood tracker app.
        \subsubsection{Setting} \label{se:setting} 
            By definition, AR integrates the virtual information within their environment \cite{azuma1997survey}.
            Consequently, in addition to affecting the user's efficiency and effectiveness in their digital data acquisition, AR interfaces influence their experience interacting and perceiving the RW and other people \cite{skarbez2021revisiting}.
            The state of the user's environment and other people, objects, and digital devices within it constitute a crucial part of the context, influencing interface design dimensions that can cause or prevent conflicts to the user's interaction and awareness of social, RW, and digital information.
            For instance, to adapt the AR content's layout and avoid interfering with a mobile user's visual attention and interpersonal interactions, contextual components providing information about semantic changes in the environment and social setting are crucial \cite{cheng2021semanticadapt, orlosky2015halo}.
            \par 
            The user category isolated the user, providing all the contextual information about them.
            The setting category describes the context without any knowledge of the user's profile and state.
            Setting is provided through three main subcategories of contextual components describing the \textit{Real-World Environment} and the \textit{Digital} and \textbf{Social} Settings.
                  
      \vspace*{5pt}
      
      \noindent  
            \textbf{Real-World Environment}\\
             Involves contextual components describing the immediate surroundings of the user and the geographical and local phenomena (\Cref{fig:contextAll}-a), regardless of their profile and state.
            \par
            \textbf{\emph{Local Environment}} is described by \textit{extracted} spatiotemporal and geospatial contextual components about the user's current location.\\
            \textit{Persistent Local Environment} includes \textit{persistent contextual components} such as geopolitical data, local climate, culture, time zone, language, health resources, and social factors.\\
            \textit{Transient Local Environment} includes \textit{Transient contextual components} such as GPS location, the local weather and time, the natural, cultural, and political phenomena, and other spatiotemporal data.
            \par
            The user's \textbf{\emph{Immediate Environment}} changes as they move throughout the day.
            The immediate environment provides transient information to describe the user's exact surroundings. 
            It includes \textit{Transient contextual components} that must be \textit{sensed} from the sensors in the user's immediate environment.
            contextual components such as temperature, lighting, noise and audio, spatial map, smoke, nearby objects, their state, and their motions are a few of these components.
                  
      \vspace*{5pt}
      
      \noindent  
            \textbf{Digital Setting}\\
            The digital setting includes contextual components that, \textit{extract} information from the available global knowledge and the immediate digital and virtual interfaces and devices without any knowledge of the user and the non-digital RW environment
            (\Cref{fig:contextAll}-b).
            \par
            \textbf{\emph{Persistent Global Knowledge}}
            includes \textit{persistent contextual components} extracted from global and common knowledge such as LLMs, search engines, public articles, social media posts, and web pages, providing information such as cultural calendars and holidays,
            \par
            \textbf{\emph{Transient Digital Setting}}
            includes \textit{Transient contextual components} that describe the current content, state, and activities of the available digital interfaces and devices within the user's immediate surroundings.
            The transient digital setting includes contextual components such as the available AR input and output modalities, the current state of its design dimensions, and the nearby devices and their states.
            For example, when driving an electric car, contextual components, such as the battery level and movement state, can affect the AR design.
                  
      \vspace*{5pt}
      
      \noindent  
            \textbf{Social Setting}\\
            Throughout the day, changes in the number of people around a user, who they are, and what they do affect the user's AR interface.
            For instance, when the user is in a public setting, concerns such as social stigma, privacy, and security can affect their preferred interaction modality.     
            As part of the Setting, Social Setting describes the other people involved in a context without any knowledge of the user.
            Social setting involves \textit{Transient contextual components} that provide information about the virtual or physical presence, state, and profile of other people, i.e., attendees, in the user context and their interactions with each other and the context(\Cref{fig:contextAll}-c).
            \par           
            \textbf{\emph{Attendee Profile}}
            includes \textit{extracted contextual components} from publicly available data or internal databases that identify the persistent profile of the present attendees. While the user profile data are persistent for each attendee, these attendee profile contextual components are transient due to changes in the presence of each specific attendee.
            The attendee profile includes each attendee's social status, occupation, relationships, values, responsibilities, etc.
            \par
            \textbf{\emph{Attendee State}}
            provides \textit{sensed inputs} that describe the transient state of each attendee.
            The attendee state includes contextual components such as face and speech detection, the physical and emotional state of each person, and their motions.
            \par
            \textbf{\emph{Attendee-Setting Interplay}}
            provides \textit{sensed} or \textit{extracted} to provide information about how the attendee's profile and state links with other people, the RW environment, and digital setting within the user's context
            The attendee SUI includes contextual components such as the personal relationships of attendees with each other and their conversation and interactions with other people and the setting.
        \subsubsection{Setting-User Interplay}\label{se:sui}
            Providing effective, efficient, and unobtrusive AR interfaces requires knowledge about the user’s needs, preferences, resources, and limitations.
            Such information can be learned from the user's background, characteristics, current state, activities, and their setting.
            However, the interplay between the User and Setting can create new needs and limitations, resolve previous ones, and lead to conflicts and associations within the context, influencing the effectiveness of the interface and how it should adapt.            
            How the user interacts with their RW and digital environment and what information they need, dictate the input and output modalities and the placements of virtual content that must be prioritized/avoided.
            The conflicts and interactions between the user and the setting are critical components of context. To account for these dynamics, previous taxonomies have incorporated \textbf{``task"} as a contextual element. However, \textbf{``task"} is often an ambiguous concept that resists quantification. In our taxonomy, Setting-User Interplay (SUI) includes Transient contextual components that can be \textit{sensed} or \textit{extracted} to provide information about how the user's profile and state links with other people and the RW environment and digital setting(\Cref{fig:contextAll}-d).
            SUI includes contextual components such as physical and digital objects, locations, dates, and people that are occluded, related to the user, or interacted with (through physical, verbal, gaze, gesture, digital, social, or emotional interactions).
            For example, SUI provides \textit{sensed contextual components} on the user's fixation point in the setting.
            Through \textit{Inferences}, this contextual component can \textit{indirectly} provide information about the people and virtual and physical objects of interest.
            Similarly, SUI contextual components from face recognition provide information for potential \textit{Inferences} that distinguish the optimal interface for formal vs. intimate social interactions.
            
\subsection{Architecture}
    \figArch  
    In this section, we propose an architecture for implementing iAR based on our proposed framework and taxonomy.
    This architecture outlines the envisioned implementation of the Context, Inference, and Adaptation Units, including input, output, and the data flow and communication mechanisms within iAR systems \Cref{fig:architecture}.
    To illustrate this, we present examples of data structures for each data type used in the architecture. These examples are for demonstration purposes only and may vary based on the specific implementation of the architecture in the real world.
    For instance, while we use a dictionary as an example data structure for contexts, depending on the system implementation, other data structures, such as a graph or a tree, may be more appropriate.
    \subsubsection{Context Unit}
    The Context Unit identifies the value of contextual components relevant to the user, their setting, and their interplay with it (SUI) from the input data specified by the user or derived from sensors and internal and external databases.
    Here we represent context as a dictionary with a key for each contextual component ($CC$).
    For example, the present context can be represented as:\\
    $Context_{(present)} = \{CC_1:val_{(CC1)}, CC_2:val_{(CC2)}, ...\}$\\
    A scenario can be represented by a sub-dictionary of context, including only the common contextual components within that scenario.
    \par
    Since the interface is part of the virtual setting, iAR directly affects the Context Unit (\Cref{fig:architecture}).
    We represent an adaptation by a heterogeneous dynamic list that includes three items: the design dimension that will be modified, the value assigned to this element by the adaptation, and a dynamic list of the entities to which the adaptation applies.
    For instance:\\
    $Adaptation_{(bodyFixed)} = [ 'FoR', 'bodyFixed',[Map, Notification Menu, ...] ]$

    \subsubsection{Inference Unit} \label{se:infer}
        The Inference Unit extracts information regarding the potential impact of specific adaptations from the context and provides it to the Adaptation Unit.
        The iAR interface deduces inferences through Rule-Based and Intelligent Inference modules.
        Here we represent an inference by a heterogeneous list comprised of two items: a dynamic list of adaptations and an impact score indicating its effectiveness:\\
        $Inference_{(bodyFixed)} = [[Adaptation_{(bodyFixed)}, Adaptation_{(transparent)}], 0.85]$
                
    \vspace*{3pt}
    
        \noindent 
        \textbf{Rule-based Inference Module:}
            The rule-based module utilizes a database of predefined Design Principles to identify inferences.
            This module uses these rules and the present context to identify the present scenarios for which a rule exists and infers the rule-based inferences.
            \par
            A scenario generalizes multiple contexts, making it easier to replicate.
            By considering a relatively small number of contexts within a scenario, patterns can be identified.
            These patterns can provide generalized Design Principles about the impact of specific adjustments in that scenario.
            For instance, a Design Principle for a walking scenario, $DP_{(walking)}$, may indicate a highly favorable impact for adapting the virtual content's frame of reference to be body-fixed, as it creates a more natural and intuitive interaction.
            Since each unique context belongs to numerous scenarios, utilizing a collection of Design Principles congruent with the present context can enhance the effectiveness of adaptations.
            \par
            A Design Principle can be represented with a heterogeneous list of two items: a scenario and a dynamic list of inferences for that scenario.
            A few examples are:\\
            $DP_{(walking)} = [Scenario_{(walking)}, [Inference_{(bodyFixed)}]]$\\
            $DP_{(street)} = [ Scenario_{(street)}, [Inference_{(multi_1)}, Inference_{(headFixed)}]]$
            \par
            Currently, the utilization of rule-based inference constitutes the primary method for the design of context-aware interfaces.
            As 
            interface designers, we manually established rules (i.e., design principles/considerations/choices) that specify the impact of particular adaptations for specific scenarios.
            We refer to these design choices, design guidelines, rules, etc. as \textit{Design Principles}.
            In this work, we distinguish iAR from context-aware AR.
            We consider \textbf{Context-aware AR} as the existing contemporary, proposed approaches where the Design Principles are the only source for providing inferences to the adaptation unit.
            For instance, consider two Design Principles for a scenario $Scenario_{(Occlusion)}$ in which the user's view of a child they are monitoring is occluded by $App_{(occluding)}$:
            $DP_{(repositionMechanism)}$ infers moving up the occluding entity, while $DP_{(OpacityMechanism)}$ infers making the occluding entity transparent \cite{davari2020occlusion}.\\
            $Adaptation_{(Move)} = [ position, position.y + 10,~App_{(occluding)}]$\\
            $Inference_{(Move)} = [ Adaptation_{(Move)}, 0.99 ]$\\
            $DP_{(repositionMechanism)} = [ Scenario_{(Occlusion)}, [Inference_{(Move)}] ]$
            $Adaptation_{(Transparent)} = [ color, color.\alpha = 0.0, App_{(occluding)}]$\\
            $Inference_{(Transparent)} = [ Adaptation_{(Transparent)}, 0.99 ]$\\ 
            $DP_{(OpacityMechanism)} = [ Scenario_{(Occlusion)}, [Inference_{(Transparent)}] ]$
            \par
            Rule-based inferences offer iAR the potential to automatically adapt even when a limited number of contextual components can be identified and extracted.
            This approach also allows iXR to utilize rules from the currently available interface design knowledge for specific scenarios.
            Additionally, the rule-based approach offers the potential for adaptability to a wide range of contexts through a small set of rules.
            However, despite its capabilities, rule-based inference is not without its limitations and challenges.
            \par
            One key limitation of this approach is the manual process of extracting and designing these Design Principles, which can be prone to human error, demanding, time-consuming, and limited to the existing knowledge.
            On the other hand, the over-generalization of the adaptations through a limited number of rules can negatively impact the performance. 
            For example, a rule mandating the Opacity of virtual content in a conversation potentially enhances the user experience by reducing distraction when the conversation primarily aims to establish social connections and the virtual content presents supplementary information.
            However, if the primary objective of the conversation is to convey information from the virtual content, this rule may degrade the interface's performance. 
            As a result, an interface that lacks the appropriate rules for the latter scenario may fail in such contexts.
            \par
            Furthermore, the effectiveness of this approach is highly dependent on the reliability and accuracy of the Design Principles. 
            For instance, incorporating a design principle on the scenario of $[Scenario_{(Y2023)} = {CC_{(year)}: 2023}]$ would impact all conceivable scenarios within the year 2023 and may prove to be detrimental if not appropriately considered.
            Another key aspect of iAR is its ability to tailor the experience to the individual user.
            However, it is difficult for the interface designer to create Design Principles that take into account the unique preferences, values, and limitations of a user who deviates from the typical, average user \cite{fischer2001user}.
            While limited user-specified rules can be introduced, the rule-based inference approach is constrained to personal experiences.
        
    \vspace*{3pt}
    
        \noindent \textbf{Intelligent Inference Module:} 
            Intelligent Inference can enhance iAR's capabilities by leveraging Machine Learning (ML) techniques to learn from the user's behavior and automatically extract inferences without any prior knowledge or rules.
            This approach allows iAR to go beyond the limitations of traditional context-aware AR interfaces.
            By observing the user's manual adaptations in different contexts, ML models can be trained to predict the impact of specific adaptations in those contexts.
            \par
            This architecture envisions the implementation of distinct ML models for each AR design dimension.
            These models generate Universal and Personalized Inferences derived from the context.
            \textit{Universal Inferences} are general inferences that can apply to numerous users, while \textit{Personalized Inferences} are tailored to the specific user's behavior.
            \par
            The Intelligent inferences generated by this unit are a collection of all these Universal and Personalized Inferences, providing a more comprehensive and accurate set of inferences.
            This approach enables iAR to adapt to the user's needs and preferences in real time, resulting in a more personalized and seamless user experience.
    \par
            \textbf{\textit{Universal Inferences Module:}}
            Universal Inferences are generated through static ML models, which are trained using batch learning on a large dataset collected from a diverse user population.
            Utilizing a large user population enables training models on a greater number of data points, resulting in a higher level of accuracy.
            These models are trained on a fixed dataset and do not update their parameters after the training process, allowing them to predict universal inferences with a relatively low computational cost. However, it should be noted that the generation of universal inferences relies heavily on the prior collection and availability of data from the implementation of AR technology.
            This poses several challenges, including the necessity of widespread adoption of AR, addressing user privacy concerns, establishing a unified infrastructure for data collection, and ensuring the availability of adequate resources for data storage and transfer.
            Further research is necessary to explore solutions to these challenges.
            One potential solution is the use of Generative Adversarial Networks (GANs) to generate training data using the available Design Principles.
            \par
            \textbf{\textit{Personalized Inference Module:}}
            Personalized Inferences are generated through online machine-learning models that continuously process the user's feedback on each adaptation in real time.
            Online learning is particularly useful in this case as it not only addresses privacy concerns but also eliminates the need for a unified infrastructure for data collection and storage, thus reducing the technological requirements for data storage and transfer.
            As new adaptations are made, these models update their parameters and adapt to the user's preferences.
            The data used for training these models includes both the user's manual interventions and the interface's automatic adaptations in each context.
            This feedback provides the online models with information on which adaptations the user accepted or rejected, enabling them to adjust their internal parameters to better suit the user's needs.
            For instance, receiving the user's manual intervention on the interface, $Adaptation_{(manual)}$, after an automatic adaptation, $Adaptation_{(automatic)}$, can serve as feedback for the machine learning model to adjust its internal parameters.
            Specifically, this feedback can reinforce a negative assessment of the effectiveness of $Adaptation_{(manual)}$, while concurrently reinforcing a positive assessment of the effectiveness of the $Adaptation_{(automatic)}$ in that particular context.
            This type of real-time learning allows the model to adapt to changing data distributions and provide personalized inferences tailored to the specific user.
            The use of transfer learning to transfer the knowledge acquired by these online learners from different users to the batch learners in the Universal Inference Module provides another potential solution to the challenges of training those static models.
    \subsubsection{Adaptation} \label{se:adapt}
        The adaptation unit utilizes the \textit{inferences} to optimize the AR interface design dimensions \cite{davari2024designspace} through two primary operational modules: Inference Consolidation and Optimization.
              
      \vspace*{3pt}
      
      \noindent  
        \textbf{Inference Consolidation Module:}
            The Inference Consolidation module is responsible for determining the actual impact of each proposed adaptation.
            In each context, multiple rule-based and intelligent inferences may provide duplicate or contradictory information about the potential impact of specific adaptations.            
            The inference consolidation module integrates various inferences involving a given adaptation and synthesizes a consensus about its potential impact.
            This module then provides the Optimization module with a single, consolidated inference for each potential adaptation.
              
      \vspace*{3pt}
      
      \noindent  \textbf{Optimization Module:}
            The optimization module combines the inferences generated by the inference consolidation module and optimizes these adjustments to select a set of adaptations that enhance the overall AR effectiveness.
              
      \vspace*{4pt}
      
            As an example, let us consider utilizing the social context-aware interface from \cite{davari2022validating} in a conversation in which the user is being asked about the weather app, currently placed in front of the interlocutor's face.
            The inference consolidation module assesses the potential impact of making the weather app fully opaque.
            In this occlusion scenario, the Design Principle, \textit{Support for viewing social cues}, infers a fully opaque weather app would have a negative effect on the user's experience.
            However, the same rule also infers that simultaneously adapting the weather app's opacity to fully opaque and moving it above the user's face would reverse this negative effect.
            On the other hand, the weather app provides socially relevant information. 
            Thus, the inference rule, Support for socially relevant information access, infers that the weather app's fully opaque opacity would have a relatively significant positive impact.
            The inference consolidation module integrates these inferences and subsequently presents the optimization module with a single comprehensive inference, which suggests a significant positive impact on the simultaneous adjustment of the weather application's opacity to fully opaque and its placement above the user's face.
            In this example, the Adaptation Unit also receives an inference suggesting a small positive impact on adapting the opacity to be fully transparent for all content, including the weather app.
            This inference, inferred from the Design Principle, \textbf{RW prioritization}, in this social scenario, will be directly provided to the optimization module without consolidation since it is the only inference on adapting the opacity to fully transparent.
            The optimization module enhances the interface's performance by rendering the weather app fully opaque and repositioning it.
\section{Experiment}\label{se:experiment4}
     Designing an effective iAR interface necessitates understanding the intricate relationship between the context and users' preferences regarding various design dimensions of the interface. This requires training data involving user adaptations across diverse contexts.
     To address \textbf{\textit{RQ3.}}, we conducted an empirical study involving a dynamic scenario with various context switches in a library.
     We observed each participant's behavior and interactions and collected data about their context and their adaptations to the AR interface.
    \subsection{Experimental Design}    
    In this experiment, participants utilized an AR interface equipped with five apps necessary for their task completion.
    Each participant encountered multiple contexts, varying in the immediate RW environment and the user's state and objective.
    They were instructed to manually adapt the availability, spatial layout, and appearance of each app at any time according to their preferences.
    \subsection{AR Interface \& Contexts} \label{se:interface}
    The AR interface provided participants with five apps to complete their tasks: a \textit{Library Map app}, displaying a picture of the library's floor plan and the call number ranges on each bookshelf, helping participants find shelves with specific call numbers;
    two \textit{Book app}s, each presenting a picture displaying a book cover along with its corresponding library call number;
    a \textit{Messaging app}, displaying simulated text messages, including a message from a friend, inviting the user to join them at a specific study area, along with an image of the area;
    and a \textit{Stock app}, allowing participants to trade simulated stocks and providing visual and auditory alerts for favorable trading opportunities.
    \par
    The session involved a context-switching scenario in which changes occurred to three contextual components:
          
      \vspace*{3pt}
      
      \noindent  
    \textit{\textbf{RW Objective:}}
     The participants were instructed to sequentially attend four objectives in the RW.
     Each participant started by searching for two books indicated within the \textit{Book apps} in the library (\textit{Locate Book1}, \textit{Locate Book2}).
     Once both of these objectives were achieved, they received a text message on the AR device. They were instructed to use an image from the \textit{Messaging app} to locate their friend in a specific study area (\textit{Locate Friend}).
     Finally, upon reaching the study area, they were asked to read a page from one of the books(\textit{Read}).
          
      \vspace*{3pt}
      
      \noindent  
    \textit{\textbf{Mobility: }}
    The participant's physical activity changed throughout the session. At times, they were \textit{Mobile}, walking within the environment, while at other times, they were \textit{Stationary}, either reading or searching through a shelf of books.
          
      \vspace*{3pt}
      
      \noindent  
    \textit{\textbf{Environment:}}
    When searching for a book, at times, the spatial characteristics of the participant's immediate environment remained \textit{Confined} within the surrounding shelves. However, at other times their immediate environment was \textit{Unconfined}, e.g., as they were walking between open areas and bookshelves.
    \\
    In each session, the participants encountered eight unique contexts: 
    \par
    $C_1$: Participants first \textit{RW Objective} involved \textit{Locating Book1}. To navigate within the library and locate the correct aisle of bookshelves, they were \textit{Mobile}, searching within an \textit{Unconfined Environment}.
    \par
    $C_2$: To \textit{Locate Book1}, participants navigated the correct shelf within an aisle of shelves, being \textit{Mobile} in a \textit{Confined Environment}.
    \par
    $C_3$: To \textit{Locate Book1}, participants were \textit{Stationary} searching for a book within a shelf in a \textit{Confined Environment}.
    \par
    To \textit{Locate Book2}, participants encountered $C_4$, $C_5$, $C_6$ respectively identical to $C_1$, $C_2$, $C_3$, with the difference in the \textit{RW Objective}.
    \par
    $C_7$: After locating both books, to \textit{Locate their Friend}, participants were \textit{Mobile} within the \textit{Unconfined Environment} of the library.
    \par
    $C_8$: In this  \textit{Stationary} context, participants \textit{Read a Book} while seated in an \textit{Unconfined Environment}.
    \\
    Throughout the session, within all these contexts, the participants were directed to utilize the \textit{Stock app} as a secondary objective to maximize their revenue (\textit{Virtual Task}).
    This \textit{Virtual Task} is to investigate any consistent pattern across all participants' adaptations to the \textit{Stock app}, i.e., the AR content utilized for a solely Virtual Task.
    \subsection{AR Adaptations} \label{se:ar_adaptations}
    To investigate the interaction between these contexts and the AR visual presentation, the AR interface provided various functionalities to adapt various design dimensions of each AR app at any time \cite{davari2024designspace}.
    \par
    \textbf{Visibility:} Each app featured a \textbf{\textit{Minimize}} button, allowing participants to remove apps from their environment or reopen them at their previous state from an app menu according to their preference. 
    \par
    \textbf{Spatial Layout:} 
    Three buttons labeled \textbf{\textit{Head-fixed}}, \textbf{\textit{Body-fixed}}, and \textbf{\textit{World-fixed}} within each app permitted participants to switch between these three fixed FoRs at any time.
    Participants had the ability to drag each app to their desired \textbf{\textit{Position}} within its FoR and adjust its \textbf{\textit{Size}} using a bimanual pinch-scale gesture.
    \par
    \textbf{Appearance:} 
    Participants used a percentage slider to adjust the \textbf{\textit{Opacity}} of each app whenever desired.
    \subsection{Apparatus}
    During the experiment, participants utilized a Microsoft HoloLens (2nd gen) AR Head-Worn Display (HWD). This wireless device features a holographic density of 47 pixels per degree and a 52-degree diagonal field of view. Our AR interface was developed using Unity and Microsoft's Mixed Reality Toolkit 2 (MRTK2).
    \figScreenshots
    \par
    We utilized pictures, as detailed in \Cref{se:interface}, for both our \textit{Library Map app} and the \textit{Book apps}.
    To create the \textit{Stock app}, we developed a custom application that featured the participant's total ``profit" along with two stock prices, each accompanied by a red ``buy" button and a green ``sell" button (See \Cref{fig:screenshots-b}).
    Participants aimed to increase their profit by purchasing stocks displayed in red text, denoting a price decrease, and selling those in green text, signaling a price increase. Price changes were announced by a notification sound, occurring randomly every 40 to 60 seconds. Consistency across participants was ensured via predefined intervals stored in a text file.
    Upon locating \textit{Book 2}, our interface issued both auditory and visual notifications to indicate a message from a "friend". Subsequently, the \textit{Messages app}, initially showing an empty chat, revealed the received text and image.
    \par
    App adaptations were exclusively enabled within the "Edit Mode," accessible via a hand-attached edit button  (Refer to \Cref{fig:screenshots-a}). Upon activation, users gained access to a hand-attached app menu, as well as UI buttons, sliders, and drag and scale gestures on individual apps, enabling app adaptations as outlined in Section \ref{se:ar_adaptations}.
    \par
    Implementing the \textit{Body-fixed} FoR in current AR devices poses challenges in accurately detecting the user's body orientation.
    In our approach, we inferred the body's forward direction based on the movements of the headset. We identified changes in body direction by comparing the current movement vector at each frame($\vec{x}_{curr}$) with the previous normalized movement vector ($\vec{x}_{prev}$). To mitigate the influence of head jitter and minor movements, we applied a threshold ($\vec{x}_{prev} \cdot \vec{x}_{curr} < 0.85$) and focused on linear movement along the floor plane, disregarding vertical Y-axis components. This method detected the user's body orientation in mobile contexts.
    To handle the body orientation changes in the stationary contexts, we introduced a "center" voice command to utilize the user's head orientation at the moment of activation as the orientation reference of body-fixed apps.
    \subsection{Procedure}
    After receiving approval from our university’s Institutional Review Board, we recruited participants from our local university. The experiment lasted 60 minutes.
    \\
    Upon arrival, participants were requested to carefully review and sign the consent form. Following this, we collected demographic information and their prior experience with AR. After this pre-study questionnaire, we provided a brief overview of the study and introduced the device, its operation, and its interaction methods.
    \\
    Before the main session, participants conducted eye calibration on the HoloLens device and underwent a training session to become acquainted with the experimental tasks and the provided AR apps and adaptations.
    They then implemented their preferred adaptations to the AR apps and proceeded to the main session to carry out the study tasks.
    All participants experienced the eight contexts in the same sequence and were encouraged to make adaptations whenever they deemed necessary.
    Moreover, they were prompted to voice their thoughts freely (think-aloud protocol), articulating reasons for making adaptations, providing specifics about each adaptation, and identifying any desired adaptations that were unavailable \cite{thingAloud}.
    After the study, all participants were interviewed regarding their overall experiences during the session.
    \subsection{Participants}
        We recruited 23 participants from our local university community. This population was between 20 and 25 years old ($M = 21.35$, $SD = 1.43$), with 21 being male and two female. Among them, 16 participants had limited AR experience, and three experienced AR more than ten times.

    \subsection{Data Collection}
        We collected information about the user's context, along with time-stamped interface adaptations and task completion time and accuracy throughout the session.
        As our contextual information, we collected CSV files logging spatial and velocity data from the user and their gaze during system use, as well as depth images and RGB videos from the forward-facing cameras on the HWD.
        The depth images had a resolution of 512x512 pixels and were recorded from the device's depth sensor at 1 frame per second (FPS).
        The videos had a resolution of 896x504 pixels and were recorded at 15 FPS from the device's external RGB camera. These videos included audio and digital holograms rendered in the headset's display.
        \par 
        Moreover, we collected each participant's time-stamped adaptations to each app's FoR (head, body, world), position (x, y, z), rotation (x, y, z), scale (x, y, z), Opacity percentage, and visibility in CSV files. 
        We also recorded participant performance, aiming to provide this dataset as preliminary training data for future machine-learning models tailored to this specific scenario.
        To record each participant's performance, we also collected CSV files logging book discovery and study area navigation speed and user's speed and accuracy in responding to stock app recommendations. 
        Metrics such as task completion time and accuracy can be utilized for the training of multi-objective optimization models.  
\section{Results}\label{se:results}
Our participants adjusted the design dimensions of 5 apps within 8 contexts. We considered each participant's final design of each app used within each context as a data point.
Three participants' data were excluded from the analysis as they were disrupted and not collected properly for the entire session.
Consequently, our data involved 800 user-specific design choices and adaptations (8 * 5 * 20).
This section analyzes these logged data and investigates patterns within user design choices across different contexts.

The \textit{Visibility} design dimension is represented by categorical data alternating between \textit{Minimized} and \textit{Visible} values.
In $660$ of the logs (82.5\%), the apps were \textit{Visible}. These apps were used to explore design choices regarding \textit{FoR}, \textit{Position}, \textit{Scale}, and \textit{Opacity}.
\par
Since Position inherently relies on the FoR, we examined this design dimension within each FoR separately. Of the $660$ $Visible$ apps, $90\%$ were Body-fixed, while only 6 were World-fixed ($\sim 1\%$).  The World-fixed apps were excluded from the analysis due to insufficient data. To avoid bias from the abundance of Body-fixed apps, Body-fixed and Head-fixed Scale and Opacity patterns were analyzed separately.
\par
In these analyses, the \textit{Opacity} values are continuous measurements from $0.15$ (Nearly invisible) to $1.0$ (fully Opaque), and the \textit{Scale} values are continuous numbers from $0.25$ to $1.5$ indicating the relative size of each app compared to a baseline app size.
Within the \textit{Head-fixed FoR}, we binned the $(x, y, z)$ app positions from the study into nine sectors, subdividing the HoloLens' FoV ($52^{\circ}$ diagonal) into three horizontal and three vertical categories (See \Cref{fig:screenshots-a}).
Within the \textit{body-fixed FoR}, 11 sectors were considered, requiring varying degrees of head-turn and physical effort to be accessed. Once the user confirms their forward-facing direction as the \textit{Center}, we subdivide the area covering the front of the user and their periphery ($< 45^{\circ}$ head turns) into three horizontal and three vertical categories, resulting in nine sectors. The \textit{Far Right} and \textit{Far Left} position sectors cover the apps positioned to the right/left side of the user's body where more extreme head-turns ($ > 45^{\circ}$) are required to access the app. The study did not observe apps in other areas, such as directly behind or above the user.
\par
We report the Likelihood Ratio for contingency analysis of our categorical design dimensions and explore correlations of \textit{Scale} and \textit{Opacity} with other categorical factors through non-parametric Kruskal-Wallis tests.
For pairwise comparisons of \textit{Scale} and \textit{Opacity} within non-binary factors, non-parametric Wilcoxon was used.

\subsection{Design Dimension Correlations}\label{se:resultDD}
    We explored various design dimensions of apps regardless of the specific context or app. This helps manually identify Design Inferences 
    Within our observations, only $\sim 18\%$ of the apps were \textit{Minimized}, and $\sim 90\%$ of the apps were \textit{Body-fixed}.
    Within both $FoRs$, nearly $85\%$ of the apps used $\geq 0.9$ opacity levels.
    \par
    In the following, we present the results of our statistical analyses of these patterns and correlations.
    \par
    Kruskal-Wallis tests revealed \textit{FoR}'s significant influence on 
    \textit{Scale} ($\chi^2{(647, 1)} = 32.83,~p < 0.0001$), showing that the \textit{Head-fixed} apps were significantly smaller ($M = 0.57,~SD~ = 0.18$) than the \textit{Body-fixed} ones ($M = 0.8,~SD~ = 0.32,~Z = -5.73,~p < 0.0001$).
    \par
\begin{figure}[htb]
        \vspace*{-5pt}
        \centering
    \includegraphics[width=1\linewidth]{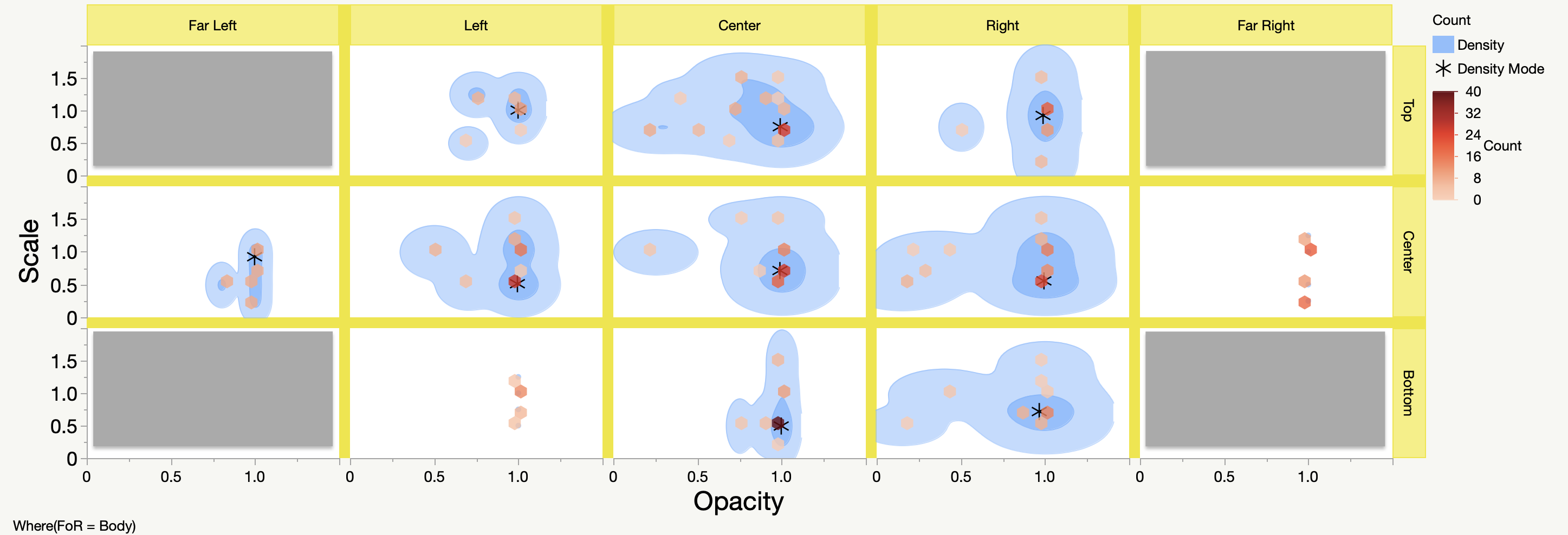}
    \vspace*{-15pt}
    \caption{HDS of Scale vs. Opacity within each Body-fixed position sector.}
    \label{fig:bodyall}
        \vspace*{-5pt}
\end{figure}

  
    Analyses of the correlations between \textit{Position}, \textit{Opacity}, and \textit{Scale} within each \textit{FoR}, indicated that Opacity correlated to position in both \textit{FoRs}, and to Scale in \textit{Head-fixed FoR}.
    \Cref{fig:bodyall} 
    depict the Highest Density Region Contours (HDS) of Scale vs. Density in each sector based on the user-specified design within the \textit{Body-fixed} FoR.

\noindent
    \textbf{Body-fixed}:\\
    $\sim 85\%$ of the \textit{Body-fixed FoR} apps were Opaque ($Opacity~\geq~0.8$) and $\sim 42\%$ were using larger scales $\geq 1$.  
    In this \textit{FoR}, \textit{Opacity} was significantly influenced by \textit{Sector} ($\chi^2{(592, 10)} = 87.4,~p < 0.0001$), \textit{Horizontal Placement} ($\chi^2{(592, 4)} = 17.63,~p \sim 0.0015$), and \textit{Vertical Placement} ($\chi^2{(592, 2)} = 24.64,~p < 0.0001$).
    \par
    All of the apps placed on the \textit{Far Right} sector and the \textit{Bottom Left} sector were fully Opaque ($M = 1.0,~SD~ = 0.0$). 
    The apps placed in the sectors \textit{above  eye-level} had \textit{Lower Opacity} ($M = 0.86,~SD~ = 0.22$), 
    compared to those that were placed \textit{at} ($M = 0.94,~SD~ = 0.18,~Z = -4.55,~p < 0.0001$) or \textit{below} ($M = 0.96,~SD~ = 0.12,~Z = -3.66,~p \sim 0.0003$) eye-level.
    \par
    \textit{Scale} was also significantly correlated to \textit{Sector} ($\chi^2{(592, 10)} = 78.94,~p < 0.0001$), \textit{Horizontal Placement} ($\chi^2{(592, 4)} = 27.81,~p < 0.0001$), and \textit{Vertical Placement} ($\chi^2{(592, 2)} = 42.88,~p < 0.0001$).
    \par
    Compared to all other vertical categories, the body-fixed apps placed in the sectors to the \textit{Left} of the user's periphery were the \textit{Largest} ($M = 0.89,~SD~ = 0.3$), and those to the \textit{Far Left} were \textit{Smallest} ($M = 0.62,~SD~ = 0.27$).
    The apps \textit{above eye-level} were also \textit{Larger} ($M = 0.94,~SD~ = 0.3$) than those placed at eye-level ($M = 0.76,~SD~ = 0.32,~Z = 5.76,~p < 0.0001$) or below it ($M = 0.74,~SD~ = 0.29,~Z = 5.92,~p < 0.0001$).
    \par
    Within the \textit{Body-fixed Center Sector}, the \textit{Opacity} and \textit{Scale} were shown to be inversely correlated by a polynomial simple linear regression ($F(65, 1) = 4.03,~p \sim 0.049$).
          
      \vspace*{1em}
      
      \noindent  \textbf{Head-fixed}:\\
    The \textit{Size} of the Head-fixed apps was inversely correlated to their \textit{Opacity}.
    $\sim 95\%$ of these apps were either placed in the \textit{Center} sector using high Opacity and scaled-down sizes ($M{Scale} = 0.52,~SD_{Scale} = 0.14,~Z = 4.61,~p < 0.0001$) or had lower Opacity and relatively larger size and were placed at the\textit{Top Left} sector ($M{Scale} = 0.83,~SD_{Scale} = 0.13,~M{Opacity} = 0.58,~SD_{Opacity} = 0.25$). $\sim 70\%$ of these apps had $Scale~\leq~0.5$, and $Scale > 1$ was never applied to any apps. 
    \par
    Our statitictical analyses reveal that \textit{Opacity} was significantly influenced by \textit{Sector} ($\chi^2{(62, 4)} = 34.8,~p < 0.0001$), \textit{Horizontal Placement} ($\chi^2{(62, 2)} = 15,~p \sim 0.0006$) , and \textit{Vertical Placement} ($\chi^2{(62, 1)} = 34.24,~p < 0.0001$).
    With one app placed at the Right side of the user's FoV (specifically at the \textit{Right Center sector}), two placed in the \textit{Top Center}, and three in the \textit{Left Center}, all the other apps were placed either in the \textit{Center Sector} ($75\%$) or the \textit{Top Left} one ($15\%$).  
    \par
    Similarly, \textit{Scale} correlated to \textit{Sector} ($\chi^2{(62, 4)} = 26.08,~p < 0.0001$), \textit{Horizontal Placement} ($\chi^2{(62, 2)} = 13.85,~p \sim 0.001$), and \textit{Vertical Placement} ($\chi^2{(62, 1)} = 25.93,~p < 0.0001$).
    \par
    Polynomial Simple Linear Regression also found a linear fit between \textit{Scale} and \textit{Opacity}, indicating their inverse correlation within the Head-fixed FoR ($F(60, 1) = 23.18,~p < 0.0001$).
\subsection{Contextual Component Correlations to Design} \label{se:resultcontext}
    The study scenario involved three contextual components and eight unique combinations of these components (i.e., Contexts). 
    Our results indicate that Context influenced when users \textit{Minimized} apps, and how they positioned their \textit{head-fixed} apps.
    The following delineates these analyses of the influence of each contextual component or context on design dimensions.
    \begin{table}[htb]
        \centering
         \vspace{-8pt}
        \resizebox{\columnwidth}{!}{
        \begin{tabular}{ccccccccc}
             & C1 & C2 & C3 & C4 & C5 & C6 & C7 & C8\\
            Visible & $91\%$ & $91\%$ & $91\%$ & $82\%$ & $82\%$ & $82\%$ & $73\%$ & $61\%$\\
            Minimized & $9\%$ & $9\%$ & $9\%$ & $18\%$ & $18\%$ & $18\%$ & $27\%$ & $39\%$ \\
        \end{tabular}}
         \vspace{-.8 em}
        \caption{User-specified app Visibility within various Contexts.}
         \label{fig:visContext}
             \vspace*{-10pt}
    \end{table}
        \par
        As shown in \Cref{fig:visContext}, \textit{Visibility} correlates to \textit{Context}.
        Our user-specified design choices indicated that the users \textit{Minimize} more apps, when \textbf{a.} in \textit{Unconfined Environments}($23\%$ of the apps compared to $14\%$ of the apps within the contexts involving \textit{Confined Environments}),
        \textbf{b.} they were \textit{Stationary} ($21\%$ of the apps compared to $15\%$ of the apps within the contexts in which they were \textit{Mobile}), and
        \textbf{c.} their objective in the RW involved \textit{Reading}($39\%$ of the apps) rather than \textit{Locating} \textit{Book1}($9\%$ of the apps), \textit{Book2}($18\%$ of the apps), or their \textit{Friend}($27\%$ of the apps).
        Likelihood Ratio tests indicated the significance of correlations between \textit{Visibility} and 
        \textbf{a.} \textit{Environment} ($\chi^2{(800, 1)} = 12.8,~p \sim 0.0003$),
        \textbf{b.} \textit{Mobility} ($\chi^2{(800, 1)} = 4.14,~p \sim 0.04$), 
        \textbf{c.} \textit{RW Objective} ($\chi^2{(800, 3)} = 48.5,~p < 0.0001$), and 
        \textbf{d.} \textit{Context}($\chi^2{(800, 7)} = 48.5,~p < 0.0001$)
        \par
        Our observations did not correlate context and contextual components to the choice of \textit{FoR} or \textit{Position}, \textit{Opacity}, and \textit{Scale} within the \textbf{Body-fixed} and \textbf{Head-fixed} \textit{FoRs}.        
        In our analyses of the \textit{Head-fixed} design dimensions, we disregard the correlations of \textit{Context} and \textit{RW Objective}. As these results are misleading considering the small number of head-fixed samples, 62 instances, leading to only 6 data points for Find Friend ($C_{7}$), and 4 for Read ($C_{8}$).
     
\subsection{App Role Correlation to Design} \label{se:resultinference}
\begin{figure*}[t!]
    \vspace*{-1.5em}
    \hspace*{-3em}
    \centering
    \includegraphics[width=1.13\linewidth]{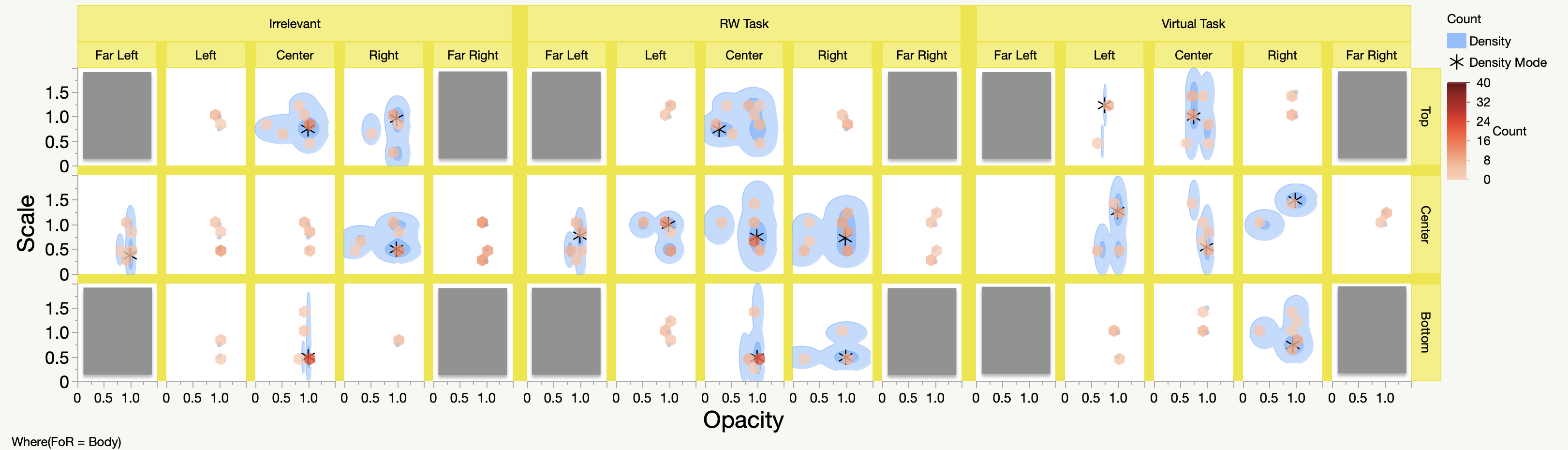}
    \setlength{\belowcaptionskip}{-18pt}
    \vspace*{-15pt}
    \caption{Influence of App Role on the user-specified design dimensions within the Body-fixed FoR}
    \label{fig:bodyInfer}
\end{figure*}
The analyses presented in \Cref{se:resultDD} and \Cref{se:resultcontext} do not consider the influence of individual apps and their content. However, depending on the user's task, the role of a particular app, and therefore, the chosen design for that app, may vary. 
\par
Considering the individual apps, our findings indicated that the Stock app was never \textit{Minimized}, and makes up $40\%$ of the uses \textit{Head-fixed} FoR.
Within $C_1$,$C_2$, and $C_3$, Book1 was never \textit{Minimized} and $20\%$ of its uses in these contexts employed \textit{Head-fixed} FoR. Conversely, in all other contexts, Book1 was \textit{Minimized} approximately $70\%$ of the time and always employed the \textit{Body-fixed} FoR when visible. 
The map app was \textit{Minimized} only in $C_8$ and consistently used \textit{Body-fixed FoR} when visible. Notably, within $c_8$, all apps except the Stock app exhibited the least usage of \textit{Body-fixed} and the highest frequency of \textit{Minimization}.
\par
    The role of each app in the user's task within various contexts can be different. 
    Depending on the \textit{Environment}, and the user's \textit{Mobility} and \textit{Objective} in the RW, an app could be \textbf{Irrelevant} in a context, \textbf{Assistive} to the user's RW task, or \textbf{Primary} to their virtual task.
    \par
    Regardless of the context, the Stock app was always \textit{Primary} for the virtual task. 
    The other four apps were \textit{Assistive} to the RW task in some contexts and \textit{Irrelevant} in others.
    In $C_8$, since the RW task of reading a physical book did not require any information from the AR interface, all four apps were \textit{Irrelevant}.
    The Messaging app was considered \textit{Irrelevant} in all contexts except for $C_7$, where it was \textit{Assistive} toward finding their friend.
    While \textit{Assistive} in finding Book1 in $C_1$,$C_2$, and $C_3$, the Book1 app was \textit{Irrelevant} in all other contexts.
    Similarly, the Book2 app was \textit{Assistive} in $C_4$, $C_5$, and $C_6$, and \textit{Irrelevant} in all other contexts.
    The map app in Contexts $C_1$, $C_4$, and $C_7$ was \textit{Assistive} for navigating through a changing and \textit{Unconfined Environment} and was \textit{Irrelevant} in the other contexts.
    \par  
    In our observations, $33\%$ of the \textit{Irrelevant} apps were \textit{Minimized}.
    Of the remaining \textit{Irrelevant} apps that were \textit{Visible}, only $5\%$ used \textit{Head-fixed}.
    The apps that were \textit{Assistive} or \textit{Primary} to the user's task were never \textit{Minimized}.
    \textit{Head-fixed FoR} was applied to $15\%$ of the time to the app that was \textit{Primary}  and $12\%$ to those that were \textit{Assistive} to the user's task.
    We explored the statistical significance of patterns involving \textit{App Role} and participants' choices of design dimensions. Our analyses indicate that \textit{App Role} significantly correlates to \textit{Visibility} ($\chi^2{(800, 2)} = 202.66,~p < 0.0001$) and choice of
    \textbf{FoR} ($\chi^2{(647, 2)} = 15.67,~p \sim 0.0004$).
    Within each \textit{FoR}, we evaluated the correlations between \textit{App Role} and \textit{Position}, \textit{Opacity}, and \textit{Scale}.
         
      \vspace*{.5em}
      
      \noindent  
    \textbf{Body-fixed}:\\
    As depicted in \Cref{fig:bodyInfer}, the \textit{Far Left} sector was never used for the \textit{Primary} app and used only for $9\%$ of the \textit{Irrelevant} and $9\%$ of the \textit{Assistive} ones.
    When \textit{Body-fixed}, compared to the other apps, the \textit{Primary} app was less frequently placed at the Vertical Center ($Primary: 37\%,~ Assistive: 63\%,~ Irrelevant: 55\%$) and more in the Top sectors ($Primary: 33\%,~ Assistive: 20\%,~ Irrelevant: 24\%$) or Bottom sectors ($Primary: 30\%,~ Assistive: 17\%,~ Irrelevant: 21\%$). These \textit{Body-fixed Position} patterns were confirmed by Likelihood Ratio tests indicating that \textit{App Role} significantly correlates to:
    \textit{Sector}($\chi^2{(585, 20)} = 89.5,~p < 0.0001$),
    \textit{Horizontal Placement}($\chi^2{(585, 8)} = 35.4,~p < 0.0001$), and
    \textit{Vertical Placement}($\chi^2{(585, 4)} = 20.5,~p \sim 0.0004$).
    \par 
    Additionally, in our observations of the \textit{Body-fixed} apps, the \textit{Primary} app was often larger and less opaque than the other apps. 
    Kruskal-Wallis tests revealed a significant influence of \textit{App Role} on both the \textit{Opacity} ($\chi^2{(2)} = 20.8,~p < 0.0001$) and the \textit{Scale} ($\chi^2{(2)} = 62.86,~p < 0.0001$) of the \textit{Body-fixed} apps.
\begin{figure}[htb]
    \centering
    \includegraphics[width=1\linewidth]{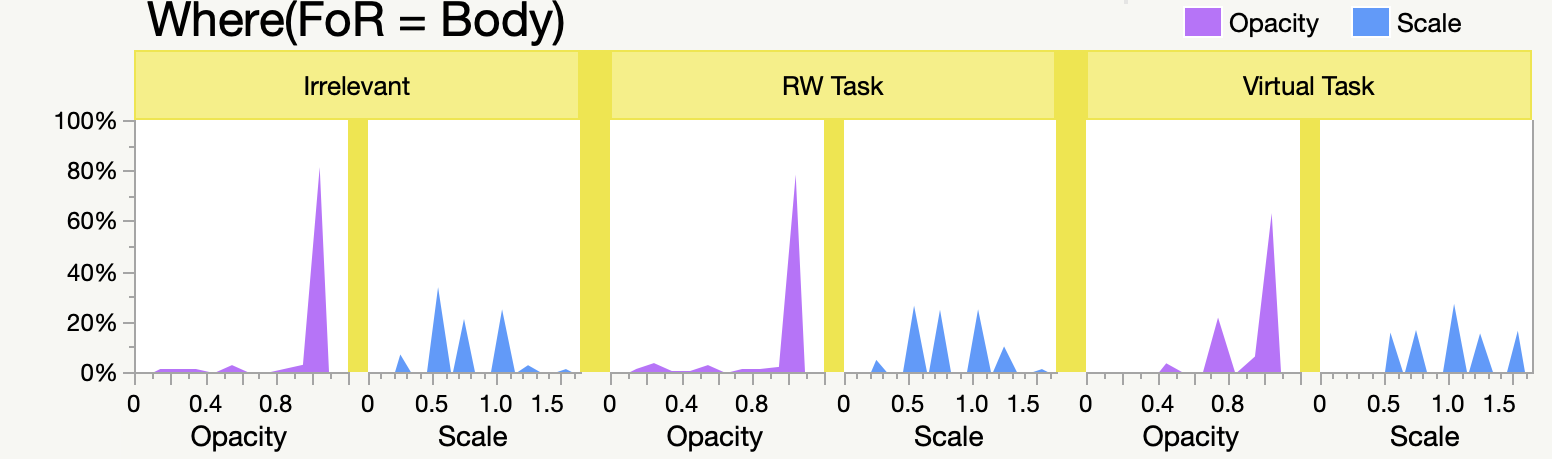}
    \includegraphics[width=1\linewidth]{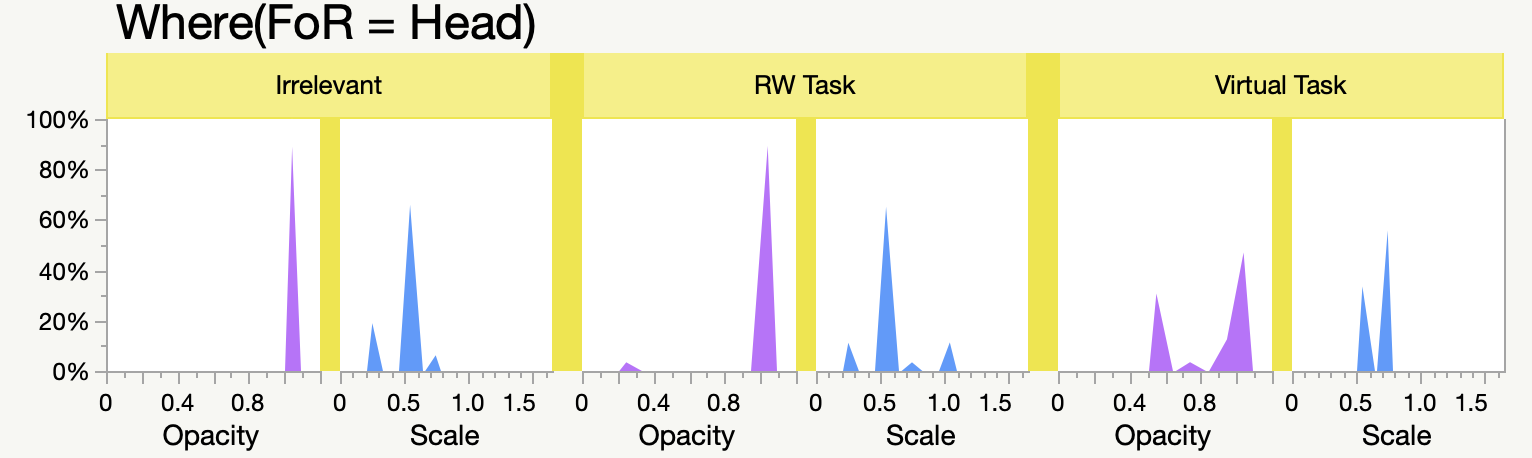}
    \caption{The histogram of the Opacity and Scale of apps with different App Roles within Body-fixed (Top) and Head-fixed (Bottom) FoR}
    \label{fig:opacityScaleRel}
\end{figure}
    \par
    Across all body-fixed apps, the \textit{Primary} app was more frequently \textit{Scaled} up to larger sizes ($M = 1,~SD~ = 0.33$) compared to the \textit{Assistive} apps ($M_{(Assistive)} = 0.78,~SD_{(Assistive)} = 0.29,~Z_{(Assistive)} = 5.4,~p_{(Assistive)} < 0.0001$) or the \textit{Irrelevant} ones ($M_{(Irrelevant)} = 0.71,~SD_{(Irrelevant)} = 0.28,~Z_{(Irrelevant)} = 7.8,~p_{(Irrelevant)} < 0.0001$).
    In this FoR, the \textit{Irrelevant} apps were also found to be often smaller than the \textit{Assistive} ones($Z = -2.52, p \sim 0.012$)(Refer to \Cref{fig:opacityScaleRel}).
         
      \vspace*{.5em}
      
      \noindent  
    \textbf{Head-fixed}:\\
    Within the \textit{Head-fixed FoR}, \textit{App Role} influenced 
    \textit{Position} ($\chi^2{(62, 8)} = 20.71,~p \sim 0.008$) and
    \textit{Vertical Placement}($\chi^2{(62, 2)} = 9.33,~p \sim 0.009$). 
    The Center Sector was used for placing the \textit{Primary} app $67\%$ of the time, $79\%$ for the \textit{Irrelevant} apps, and $83\%$ for the \textit{Assistive} ones.
    While the \textit{Primary} app was placed in the \textit{Top Left} sector $25\%$ of the time, the \textit{Irrelevant} apps never used this sector, and only $13\%$ of the \textit{Assistive} ones used it.
    \par
    Kruskal-Wallis tests revealed a significant influnce of \textit{App Role} on \textit{Opacity} ($\chi^2{(2)} = 18.4,~p < 0.0001$) and \textit{Scale} ($\chi^2{(2)} = 15.16,~p \sim 0.0005$).
    Pairwise comparisons using the Wilcoxon method indicated that the \textit{Primary} app often had \textit{Larger Scale} ($M_{(Scale, Primary)} = 0.66,~SD_{(Scale, Primary)} = 0.12$) and \textit{Lower Opacity} ($~M_{(Opacity, Primary)} = 0.81,~SD_{(Opacity, Primary)} = 0.23$) compared to the Scale ($M_{(Scale, Assistive)} = 0.54,~SD_{(Scale, Assistive)} = 0.2,~Z_{(Scale, Assistive)} = 2.81,~p_{(Scale, Assistive)} \sim 0.0048$) and Opacity ($~M_{(Opacity, Assistive)} = 0.97,~SD_{(Opacity, Assistive)} = 0.15,~Z_{(Opacity, Assistive)} = -3.32,~p_{(Opacity, Assistive)} \sim 0.0009$) of the \textit{Assistive} ones.
    Additionally, the \textit{Irrelevant} apps were always fully Opaque ($Opacity_{(Irrelevant)} = 1$) when \textit{Head-fixed} and Scaled down to smaller sizes ($M_{(Scale, Irrelevant)} = 0.46,~SD_{(Scale, Irrelevant)} = 0.13$). Therefore, compared to the these \textit{Irrelevant} apps, the \textit{Primary} app was significantly \textit{Larger} with \textit{Lower Opacity} ($~Z_{(Scale, Irrelevant)} = 3.59,~p_{(Scale, Irrelevant)} \sim 0.0003$, $~Z_{(Opacity, Irrelevant)} = -3.09,~p_{(Opacity, Irrelevant)} \sim 0.002$).
    \par
    In \Cref{se:resultDD}, we found an inverse correlation between the \textit{Scale} and \textit{Opacity} of the \textit{head-fixed} apps. 
    Further polynomial simple linear regressions on each App Role found that this inverse correlation does not apply to head-fixed apps that are \textit{Irrelevant}. 
\section{Discussion}\label{se:discussion}
Our study explored the influence of various contextual components on the design dimensions of various AR apps, revealing notable patterns and correlations, and highlighting the complexity of designing context-aware AR and the significance of our framework in enabling iAR.
The patterns observed across all apps and contexts suggest:
\begin{itemize}
\setlength\itemsep{-.4em}
    \item Body-fixed is more commonly used than head-fixed.
    \item World-fixed is rarely used in a mobile scenario.
    \item The scale and opacity of head-fixed apps are inversely correlated.
    \item Head-fixed apps are smaller than body-fixed ones.
    \item Head-fixed apps are typically positioned near the center of the FoV or top left corner.
    \item Placement below eye level is not preferred for head-fixed.
    \item Head-fixed apps near the center are usually small.
\end{itemize}
These findings can suggest several potential design principles for future AR interfaces.
However, our results indicate inconsistencies in some design principles when applied to specific contexts or apps. For instance, the inverse correlation between the scale and opacity of head-fixed apps was 
significantly influenced by App Role and various contextual components, highlighting the necessity of context awareness.
\par
Furthermore, the impact of App Role on AR design suggests:
\begin{itemize}
\setlength\itemsep{-0.4em}
    \item Relevant (\textit{Assitive} or \textit{Primary}) apps are always visible.
    \item Relevant body-fixed apps are preferably placed at eye level.
    \item The apps primary to the user task are larger with lower opacity.
    \item Irrelevant apps rarely use head-fixed FoR.
    \item Irrelevant body-fixed apps are rarely placed in the Center sector.
    \item Irrelevant apps are smaller.
\end{itemize}
These observations underscore the importance of App Role in determining the optimal design and placement of AR apps.
\par
Our results suggest that: 1) \textit{RW Objective} significantly influenced the position and visibility of head-fixed applications; 2) \textit{Mobility} and \textit{Environment} impact AR interface management, showing that users tended to minimize more apps when stationary or in an unconfined environment.
The limited correlations between design dimensions and contextual components may result from the varying importance of different contextual components. Participants adapted interfaces carefully when the \textit{RW Objective} changed but less so across different contexts with the same \textit{RW Objective}, explaining the higher similarities across various \textit{Environment} and \textit{Mobility}. Additionally, participants rarely adjusted their interfaces between consecutive \textit{Mobile} contexts, contributing to inconsistencies in the same \textit{Environment}.
\par
Additionally, all contexts in this experiment were within the same scenario. Common contextual components and implicit contextual factors across these contexts may have overshadowed the influence of the evaluated components. For instance, in this scenario, the user's task involves traveling to various locations within the library. Consequently, even in contexts where the user is \textit{Stationary}, this contextual factor remains constant. Such implicit contextual information could be the primary reason for not using world-fixed FoR, leading to our observation that \textit{Body-fixed FoR} is more commonly used. Future work exploring these common implicit and explicit contextual factors can provide evidence for the independence of design dimension correlations from specific common contextual components.
\par
While our study provides valuable insights into AR interface design, it also highlights the complexities and need for further research to develop robust context-aware AR.
Identifying generalizable patterns for adaptations emphasizes the need for robust factorial analyses and logistic regression models to understand the interaction effects of all independent variables, i.e., contextual components. 
Even with three contextual components within this study, data from all 16 possible combinations would result in more meaningful conclusions. 
Exploring a large number of contexts requires extensive data collected from all these unique contexts.
This poses not just a limitation of our work but a significant practical challenge in designing iAR.
\par
As proposed in our framework, inferences allow for the integration of various contextual components, including AR apps and their design dimensions, to identify generalizable rules applicable to different contexts.
This study underscores the significance of \textit{Inferences} such as App Role in determining the optimal AR design.
\par
Our framework proposes to use both rules and machine learning to make various inferences relevant to specific contexts and optimize over conflicting ones to provide the optimal AR interface.
The rule-based inference module in our architecture can utilize findings from this and similar studies as rules to deduce various inferences relevant to specific contexts.
On the other hand, the intelligent inference module in our architecture allows for intelligently handling these inferences and further refining the interface.
\section{Conclusions}
To adapt the presentation and interaction of virtual content and improve effectiveness, an iAR interface must collect and meaningfully integrate all relevant context information. We introduced a framework for context-aware inferences and adaptation within iAR to optimize AR effectiveness (see \Cref{se:architecture}). This framework defines a taxonomy of contextual components, including user, setting, and SUI knowledge.
\par
We presented an architecture that utilizes these contextual components to automatically optimize AR adaptations. The rule-based and intelligent inference modules in our architecture enable the AR interface to infer the impact of certain adaptations on performance. These modules cover existing context-aware designs and extend beyond them by learning to personalize and adapt to user behavior. The Adaptation Unit consolidates and optimizes inferences, enhancing effectiveness.
\par
Our experiment was conducted across multiple contexts, with varying \textit{Mobility}, \textit{Environment}, and \textit{RW Objectives} in a library. Our analysis of user-specified adaptations to the design dimensions of various apps in this experiment yielded several key insights and potential design principles. We identified that opacity is often reduced to mitigate the occlusion of the real world, especially for larger head-fixed or centrally placed body-fixed apps. Other principles include a preference for body-fixed FoR, an inverse correlation between the scale and opacity of head-fixed apps, and specific placement preferences for head-fixed apps. We also discovered that irrelevant apps were smaller and the primary app was always visible, larger, and with lower opacity.
\par
Our research enhances the understanding of how contextual components, including AR design dimensions, impact AR interface design. We present a framework for optimizing AR experiences based on context-specific inferences, paving the way for future iAR research to develop more adaptive, user-friendly interfaces.

\acknowledgments{
	Special thanks to Missie Smith and Jerald Thomas for their contributions to the direction of this work.
}
\bibliographystyle{abbrv-doi-hyperref-narrow}
\bibliography{main.bib}

\begin{tabular}{ m{0.2\textwidth} m{0.75\textwidth} }
    \begin{minipage}{0.12\textwidth}
        \includegraphics[width=\linewidth]{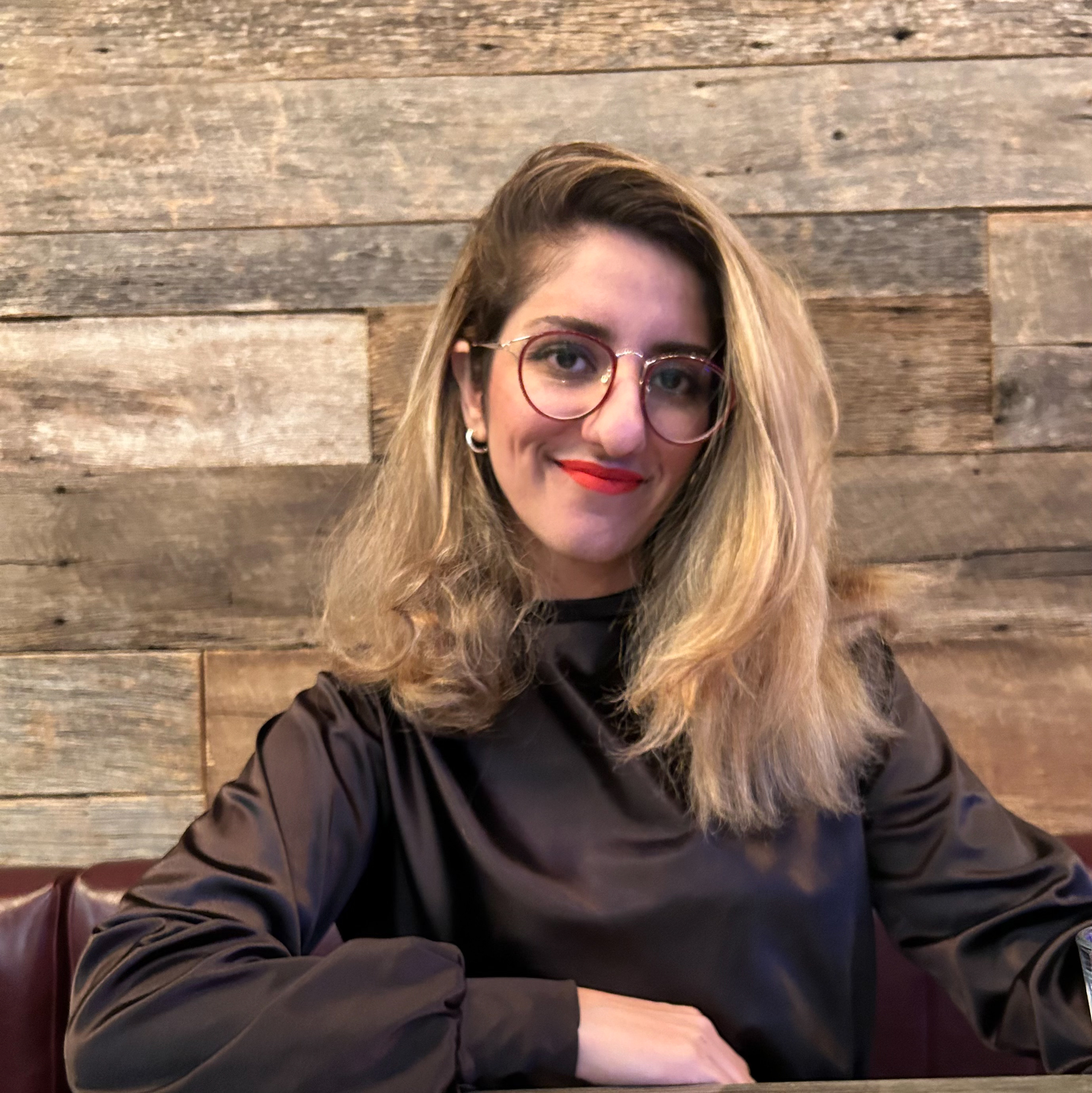}
    \end{minipage} & \hspace{-5em}
    \begin{minipage}{0.33\textwidth}
        \textbf{Shakiba Davari} is a recent Ph.D. graduate from the 3D Interaction (3DI) Group at Virginia Tech. Her research focuses on methodological guidelines for intelligent XR(iXR) and developing context-aware interfaces that address AR challenges such as occlusion and social intrusiveness.\\
    \end{minipage} \\
    \begin{minipage}{0.12\textwidth}
        \includegraphics[width=\linewidth]{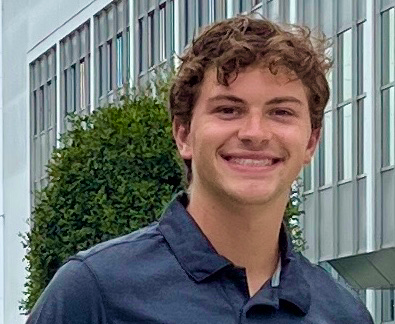}
    \end{minipage} & \hspace{-5em}
    \begin{minipage}{0.33\textwidth}
        \textbf{Daniel Stover} is a recent M.Sc. graduate from the 3DI Group and Virginia Tech's ECE Department. His thesis is focused on general-purpose task guidance in AR using machine learning, contributing to the research area of context-aware AR. \\
    \end{minipage}\\
    \begin{minipage}{0.12\textwidth}
        \includegraphics[width=\linewidth]{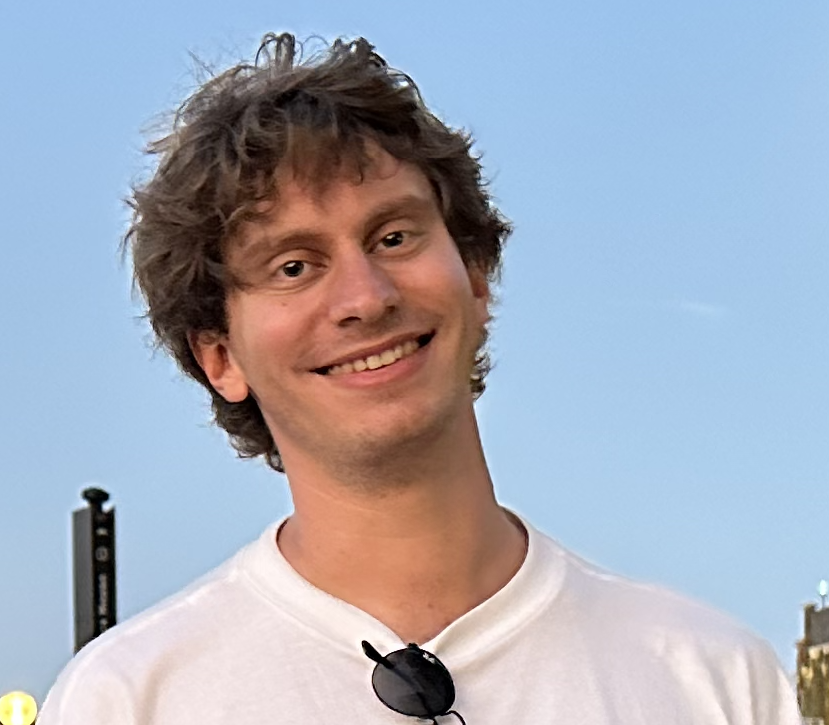}
    \end{minipage} & \hspace{-5em}
    \begin{minipage}{0.33\textwidth}
        \textbf{Alexander Giovannelli} is a Ph.D. student and member of the 3DI Group at Virginia Tech. His research focuses on communication and collaboration within immersive experiences using VR/AR technologies. \\
    \end{minipage}\\
    \begin{minipage}{0.12\textwidth}
        \includegraphics[width=\linewidth]{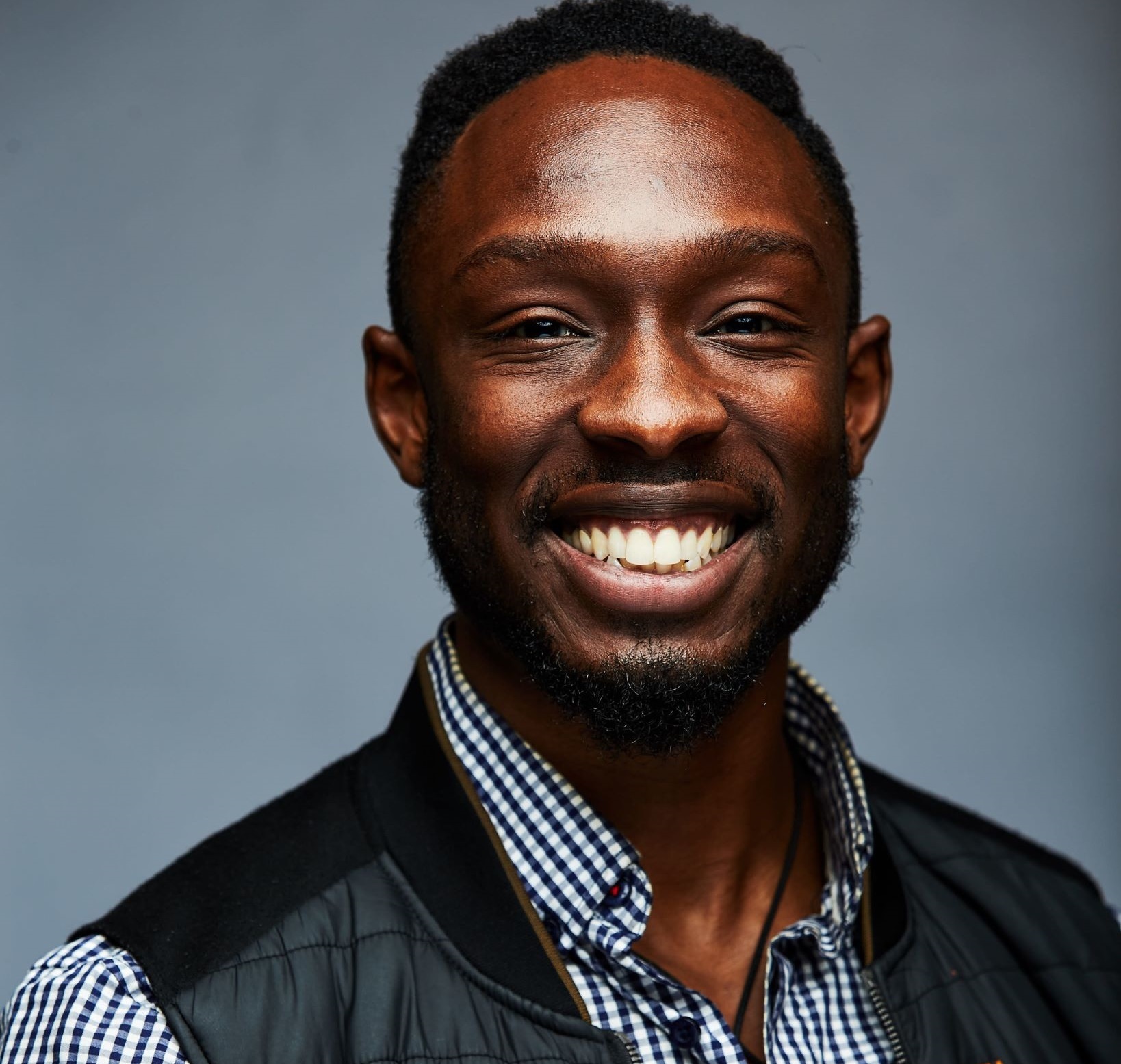}
    \end{minipage} & \hspace{-5em}
    \begin{minipage}{0.33\textwidth}
        \textbf{Cory Ilo} is a Ph.D. student and member of the 3DI Group at Virginia Tech. His research focuses on the usages of gaze data to infer user intent to better enable context-aware VR/AR applications.\\
    \end{minipage}\\
    \begin{minipage}{0.12\textwidth}
        \includegraphics[width=\linewidth]{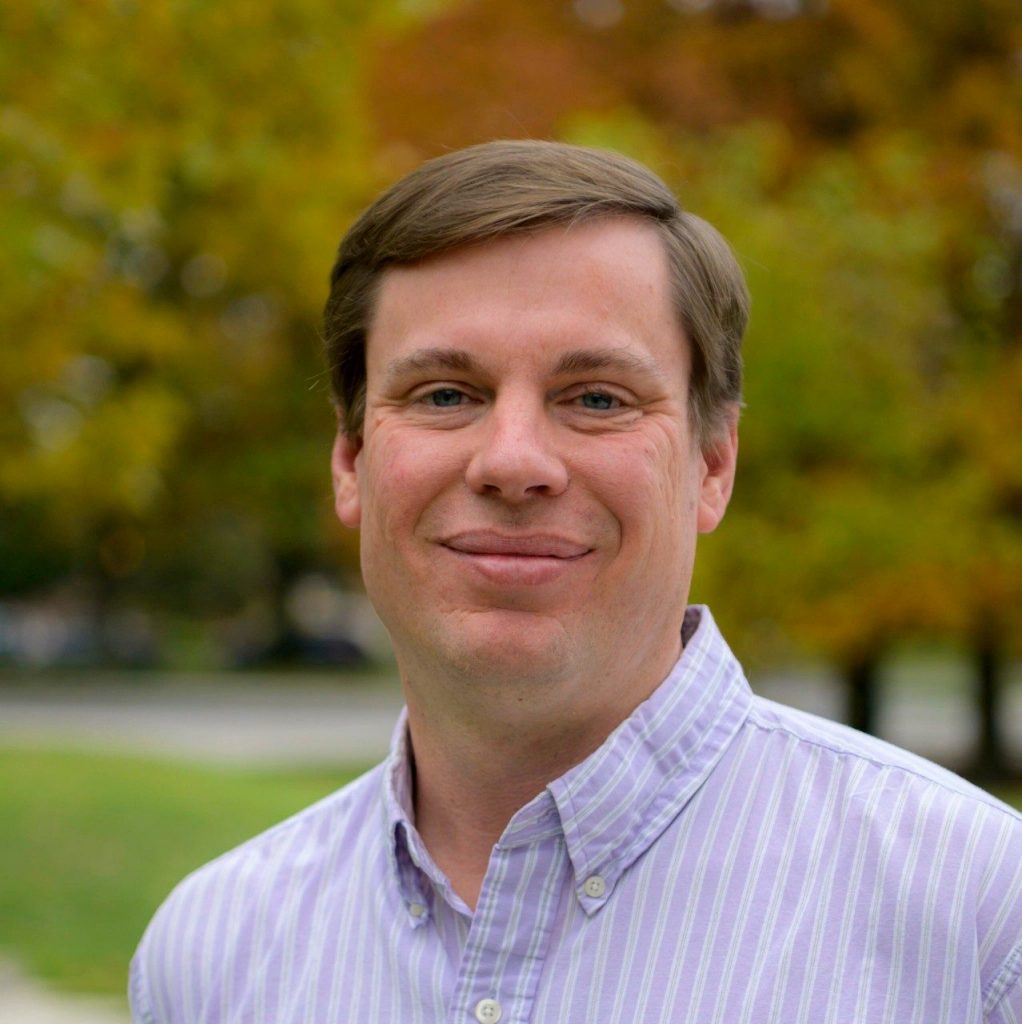}
    \end{minipage} & \hspace{-5em}
    \begin{minipage}{0.33\textwidth}
        \textbf{Doug A. Bowman} is the Frank J. Maher Professor of Computer Science and Director of the Center for Human-Computer Interaction at Virginia Tech. He is the principal investigator of the 3DI Group, focusing on the topics of three-dimensional user interfaces, VR/AR user experience, and the benefits of immersion in virtual environments \\
    \end{minipage}
\end{tabular}
\end{document}